\documentclass[notitlepage, oneside]{book}

\usepackage{graphicx}
\usepackage{latexsym}
\usepackage{amsmath}

\begin{document}

\setcounter{chapter}{11}
\chapter[]{THERMODYNAMICS, KINETICS AND FRAGILITY OF BULK METALLIC GLASS FORMING LIQUIDS }

\textbf{{\large Ralf Busch\footnote{r.busch@mx.uni-saarland.de}, Zach Evenson,  Isabella Gallino and Shuai Wei} \\ \\Saarland University, Department of Materials Science and Engineering, Campus C6.3, 66123 Saarbr\"ucken, Germany\\}




\date{\today}
\noindent{\em This review deals with the kinetic and thermodynamic fragility of bulk metallic glass forming liquids. The experimental methods to determine the kinetic fragility, relaxation behavior and thermodynamic functions of undercooled metallic liquids are introduced. Existing data are assessed and discussed using the Vogel-Fulcher-Tammann equation and in the frameworks of the Adam-Gibbs as well as the Cohen-Turnbull free volume approach. In contrast to pure metals and most non glass forming alloys, bulk glass formers are moderately strong liquids. In general the fragility parameter $D^{*} $ increases with the complexity of the alloy with differences between the alloy families, e.g. noble-metal based alloys  being more fragile than Zr-based alloys. At least some bulk metallic glass forming liquids, such as Vitreloy 1, undergo transitions from a fragile state at high temperatures to a strong state at low temperatures with indications that in Zr-based alloys this behavior is a common phenomenon. }




\section{INTRODUCTION}

The slow kinetics in bulk metallic glass (BMG) forming liquids is the main contributing factor to their high glass forming ability (GFA). Recent studies of the melt viscosity of multi-component, Zr-based BMG alloys have produced a picture of a very viscous liquid, in which the sluggish kinetics impedes the nucleation and growth of crystals [1-2]. The high viscosity originates from the large size mismatches of the liquid's many atomic species, leading to less free-volume available for viscous flow. In fact, it has been shown that the GFA of Zr-based alloys increases with an increasing number of components [3]. The underlying reason for this behavior is most likely that, with increasing number of differently sized atomic species, it becomes possible to produce higher and higher density liquids. It also makes the liquid more viscous and thus stronger in the framework of the fragility concept [4].

The formation of a glass during undercooling from the liquid is associated with the "freezing in" of a certain amount of excess free volume as the liquid falls out of equilibrium at the glass transition temperature, $ T_{g} $. A good conceptualization of the free volume of a melt, as the glass transition is approached, is given by understanding the kinetic slowdown in terms of viscosity or relaxation time. The change in viscosity or relaxation time with temperature reflects an intrinsic property of glass-forming liquids known as fragility. Glass formers that show very little change in their equilibrium viscosity or relaxation time as $ T_{g} $ is approached are defined as being kinetically "strong", i.e. exhibiting an Arrhenius dependence of the viscosity or relaxation time on temperature, where substances whose equilibrium viscosities vary much greater with temperature are classified as "fragile" [5, 6]. The temperature dependence of the equilibrium viscosity can be described with the empirical Vogel-Fulcher-Tammann (VFT) equation [7-9]

\begin{equation}
f=f_{0}exp \left(\frac{D^{*}T_{0}}{T-T_{0}} \right),
\label{Eq1}
\end{equation}
where $ f $ represents viscosity or relaxation time and $ f_{0} $ the pre-exponential factors, $\eta_{0}$ or $\tau_{0}$, which give the theoretical infinite-temperature limits for viscosity and relaxation time, respectively. The parameter $ D^{*} $ is the kinetic fragility of the material; the most fragile glass-formers have a fragility of around 2, whereas the strongest are on the order of 100. The VFT temperature, $ T_{0} $, is the temperature at which the barriers with respect to flow would approach infinity [6]. The pre-exponential factor, $ \eta_{0} $, is kept fixed at a value of $ 4 \times 10^{-5} $ Pa s, according to the relation $\eta_{0} = hN_{A}/v_{m}$, where $ h $ is Planck's constant; $ N_{A} $ is Avogadro's number and $ v_{m} $ is the atomic volume [10]. This is important, when data exist only in a small temperature range leading to a reliable VFT fit. It is justified since in the Angell plot the extrapolations of all liquids with different fragilities  meet at infinite temperature and a viscosity of $\eta_{0}$. 

A phenomenological model of the equilibrium viscosity of glass forming substances was formulated in terms of the free volume by Doolittle [11]

\begin{equation}
\eta=\eta_{0}exp\left(\frac{bv_{m}}{v_{f}}\right)
\label{Eq2}
\end{equation}
where $ v_{f} $  is the average free volume per atom of the equilibrium liquid and the parameter $ b $ is a material specific constant of order unity. The term $ bv_{m} $ represents the critical volume necessary for viscous flow. In our studies we define the free volume as in Ref. [12]; namely, it is the difference between the specific volume and occupied volume at a given temperature. The model by Cohen and Turnbull assumes a linear relation between the free volume and temperature [12]

\begin{equation}
v_{f}=v_{m}\alpha(T-T_{0}),
\label{Eq3}
\end{equation}
where $ \alpha_{f} $ can be approximated as the difference between the volumetric thermal expansion coefficients of the liquid and the glass, $ \alpha_{f} = \alpha_{liq}-\alpha_{glass}$ [13, 14]. In this model of the free volume, viscous flow occurs as a result of random density fluctuations that allow for diffusion of individual atoms without change to the local free energy [15]. In other words, viscous flow is attributed not to energy barriers, but rather to the redistribution of free volume. Assuming now that $ T_{0} $ is the temperature, at which the free volume of the equilibrium liquid would vanish and viscous flow no longer be possible, it becomes immediately apparent that by substituting Eq. (\ref{Eq3}) for the free volume in Eq. (\ref{Eq2}), the VFT equation (Eq. (\ref{Eq1})) is recovered with the relation $ \alpha_{f} = b/(D^{*}T_{0}) $. In an extended model of the free volume by Cohen and Grest [16] the metastable, equilibrium liquid is partitioned into cells, whose free energy is a function of the cell volume. Each cell behaves then either liquid-like – capable of diffusive motion, or solid-like – capable of only oscillatory motion. Taking the Cohen and Grest expression for the free volume,

\begin{equation}
v_{f}=\frac{k}{2\varsigma_{0}}\left(T-T_{q}+\sqrt{(T-T_{q})^2+\frac{4v_{a}\varsigma_{0}}{k}T} \right)
\label{Eq4}
\end{equation}
and inserting it into Eq. (\ref{Eq2}) yields the parameters $ bv_{m}\varsigma_{0}/k $ , $ T_{q} $ and $ 4v_{a}\varsigma_{0}/k $. In this newer model of the free volume, $ v_{f} $ does not vanish at $ T_{0} $. Instead, the free volume remains greater than zero at all temperatures and only vanishes when $ T = 0 $. The viscosity, therefore, would not diverge and remain well defined for all temperatures.

Another phenomenological model of the equilibrium viscosity, based on the thermodynamic functions of the undercooled liquid, is the Adam-Gibbs entropy model for viscous flow [17]

\begin{equation}
\eta=\eta_{0}exp\left(\frac{C}{TS_{c}(T)}\right),
\label{Eq5}
\end{equation}
where $ S_{c}(T) $ is the configurational part of the entropy of the equilibrium liquid and the parameter $ C $ can be understood as a free energy barrier per particle for cooperative rearrangements. The function $ S_{c}(T) $ can be calculated from the experimentally determined thermodynamic functions of the material as

\begin{equation}
S_{c}(T)=S_{c}(T_{m}^{*})-\int_{T}^{T_{m}^{*}}\frac{\Delta C^{l-x}_{p}(T^{\prime})}{T^{\prime}}dT^{\prime},
\label{Eq6}
\end{equation}
leaving the parameters $ C $ and $ S_{c}(T_{m}^{*}) $ to be determined through fitting of the experimental data, where $ T_{m}^{*} $ is a scaling parameter chosen as the temperature, for which the viscosity of the melt has a value of 1 Pa s [18]. It is assumed here that the vibrational contribution to the entropy for both the undercooled melt and the crystal are similar. Hence, the configurational part of the entropy decreases during undercooling with the same rate as the entropy difference between the liquid and the crystal, where $\Delta C_{p}^{l-x}$ is the difference in the specific heat capacities of the liquid and the crystal. By comparing Eqs. (\ref{Eq2}) and (\ref{Eq5}) it is now possible to express the relative free volume of the equilibrium liquid in terms of its configurational entropy as
\begin{equation}
\frac{v_{f}}{v_{m}}=\frac{bTS_{c}(T)}{C}.
\label{Eq7}
\end{equation}

The free volume has been recently investigated in various bulk metallic glass (BMG) forming systems through, for example, direct density measurements [19-22], viscosity studies [13, 23-26] and positron annihilation lifetime measurements [27]. Van den Beukel and Sietsma proposed a method for quantifying the free volume in terms of enthalpy as measured using Differential Scanning Calorimetry (DSC) [28]. This method has since been employed by several researchers, using enthalpy relaxation to examine the free volume of BMGs at temperatures below the glass transition [20, 22, 29-35].

\section{EXPERIMENTAL DETERMINATION OF THE THERMODYNAMICS AND KINETICS IN BULK METALLIC GLASS FORMERS}

BMG specimens are usually produced by first making a master alloy. This is done by melting the pure elements in an arc melter, if the alloy contains refractory elements such as Nb. If there are no refractory elements involved like in the case of most noble metal based BMG the alloys are melted inductively. To produce amorphous samples the master alloys are typically cast into water cooled, oxygen free copper molds to obtain rods with diameters of typically 2-5 mm or plates with a geometry of typically $ 3 \times 13 \times 34 $ mm. X-ray diffraction (XRD) is used to prove the glassy state of the alloys. The composition of the alloys is given in atom percent like, e.g. Pt$ _{60} $Cu$ _{16} $Co$ _{2} $P$ _{2} $ . Therefore enthalpies and entropies are stated in kJ g-atom$ ^{-1} $ and J g-atom$ ^{-1} $ K$ ^{-1} $, respectively. 

\subsection{THERMODYNAMICS}
The thermodynamic functions of the alloys are determined as a function of temperature. The heats of fusion and enthalpies of crystallization are measured by Differential Thermal Analysis (DTA) and DSC, respectively. The absolute specific heat capacity, cp, of the samples can be determined in a power compensated DSC of the Perkin Elmer series with great accuracy up to a temperature of 1000 K. This works on heating and cooling in reference to the specific heat capacity of a standard sapphire using the so-called 'step method' [36]. This method consists of heating the sample in steps of 20 K with a rate of 0.33 K/s and annealing isothermally for 120 s during each step. This resulted in steps in heat flux, from which the specific heat capacity of the sample can be calculated. For each step the heat flux change is:

\begin{equation}
\dot{Q}=\left(\frac{\partial Q}{\partial t}\right)_{\dot{T}\neq 0}-\left(\frac{\partial Q}{\partial t}\right)_{\dot{T}=0}=c\times \frac{dT}{dt}.
\label{Eq8}
\end{equation}

The term $ (\partial Q/\partial t)_{\dot{T}\neq 0}$ is the required power to heat sample and pan with a constant heating rate, $ (\partial Q/\partial t)_{\dot{T}=0}$ is the required power to maintain a constant temperature of sample and pan, and c is the sum of the heat capacity of the sample and sample pan. The specific heat capacity of the sample,$  c_{p}(T) $ sample, as a function of temperature is calculated by

\begin{equation}
c_{p}(T)_{sample}=\frac{\dot{Q}_{sample}-\dot{Q}_{pan}}{\dot{Q}_{sapphire}-\dot{Q}_{pan}}\times \frac{m_{sapphire}\cdot \mu_{sample}}{m_{sample}\cdot \mu_{sapphire}}\times c_{p}(T)_{sapphire},
\label{Eq9}
\end{equation}
where $ m $ is the mass, $ \mu $ the molecular weight, and $ c_{p}(T)_{sapphire} $ the standard specific heat capacity of the sapphire, tabulated in ASTM charts.

\begin{figure}[ht]
\begin{center}
\includegraphics[width=3in]{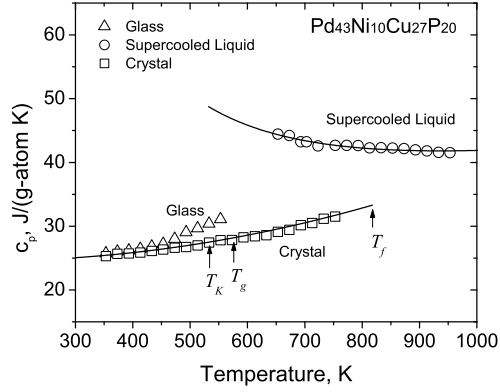}
\caption{Specific heat capacity data for Pd$ _{43} $Ni$ _{10} $Cu$ _{27} $P$ _{20} $ as a function of temperature. The symbols are data for the amorphous alloy (triangles), the crystalline solid (squares), and the liquid state (circles).The data were measured in steps in reference to sapphire. The error bar has the size of the symbols. The continuous curves represent the fits to Eqs. (\ref{Eq10}). $ T_{K} $, $ T_{g} $, and $ T_{f} $  are the Kauzmann temperature (532 K), the calorimetric glass transition temperature for heating rates of 0.33 K/s (582 K), and the calorimetric fusion peak temperature (818 K), respectively [18, 37].}
\label{Fig1}
\vspace*{-12pt}\end{center}
\end{figure}

Figure \ref{Fig1} shows the measured cp for the Pd$ _{43} $Ni$ _{10} $Cu$ _{27} $P$ _{20} $ alloy. The $ c_{p} $ of the glass and supercooled liquid was measured with the step method from 323 K to 600 K. The $ c_{p} $-measurement for the crystalline state was subsequently performed from 323 K up to 770 K. The low liquidus temperature of 850 K of this alloy permits to measure the absolute specific heat capacity of the melt using a Perkin Elmer DSC. For this purpose undercooling experiments of the liquid were performed from 953 K down to 650 K. After melting in the DSC, the sample is undercooled to an assigned temperature and then held isothermally. The DSC output shows a step in the heat flux  that is proportional to the heat capacity as in Eq. (\ref{Eq8}). The same procedure was repeated on a standard sapphire and on the empty pan in order to calculate the cp of the sample with Eq. (\ref{Eq9}).  Moreover, the isothermal anneals of the undercooled Pd$ _{43} $Ni$ _{10} $Cu$ _{27} $P$ _{20} $ melts were held long enough to detect the crystallization event. At each isothermal temperature the enthalpy of crystallization, $ \Delta H_{x} $, was measured by integration of the calorimetric crystallization peak (see Ref. 18 for details).  The calculation of the integral term in Eqs. (\ref{Eq6}, \ref{Eq11}, \ref{Eq12}, \ref{Eq13}) requires fitting calorimetric specific heat capacity data of the liquid and of the crystal to the equations below:

\begin{align}
c_{p}^{l}(T)=3R+a\times T+b\times T^{-2};\nonumber \\ 
c_{p}^{x}(T)=3R+c\times T+d\times T^{2};\nonumber \\
 \Delta c_{p}^{l-x}(T)=c_{p}^{l}(T)-c_{p}^{x}(T).
 \label{Eq10}
\end{align}

The constants $ a $, $b$, $c$, and $d$ are fitting constants, $R$ is the universal gas constant, and $T$ is the absolute temperature. When possible, the specific heat capacity curve of the liquid, $ c_{p}^{l} $, was optimized by considering, during the fitting procedure, the value of the area that lies between the $ c_{p} $ of the liquid and that of the crystal in the temperature range from the crystallization and the melting event. This area is equal to 
\begin{equation}
\int_{T}^{T_{f}}\Delta c_{p}^{l-x}(T^{\prime})dT^{\prime}=\Delta H_{f}-\Delta H_{x},
\label{Eq11}
\end{equation}
where $ \Delta H_{f} $ and $ \Delta H_{x} $ are the experimental heat of fusion and heat of crystallization, respectively. $ T_{x} $ is selected as the calorimetric onset of crystallization for heating rates of 0.333 K/s, and $ T_{f} $ is the peak temperature of the calorimetric melting signal. 

\begin{figure}[ht]
\begin{center}
\includegraphics[width=3in]{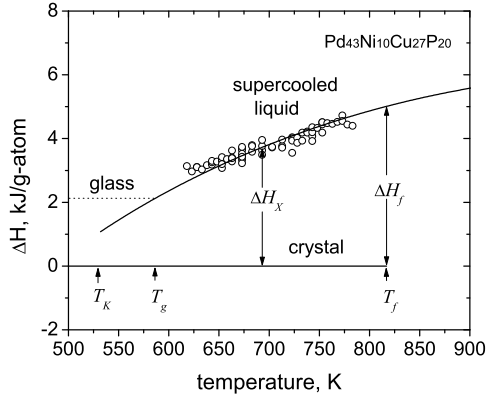}
\caption{Enthalpy difference of the undercooled liquid Pd$ _{43} $Ni$ _{10} $Cu$ _{27} $P$ _{20} $  with respect to the crystal as a function of temperature. The continuous line represent the calculated function with Eq. (\ref{Eq12}). The circles represent the enthalpy of crystallization values directly measured from isothermal undercooling experiments in DSC. $ T_{K} $, $ T_{g} $, and $ T_{f} $  are the Kauzmann temperature (532 K), the calorimetric glass transition temperature for heating rates of 0.33 K/s (582 K) and the calorimetric fusion peak temperature (818 K), respectively [18, 37]. $ \Delta H_{f} $  and $ \Delta H_{x} $  are the measured heat of fusion (5 kJ/g-atom) and heat of crystallization, respectively.}
\label{Fig2}
\vspace*{-12pt}\end{center}
\end{figure}

For Pd$ _{43} $Ni$ _{10} $Cu$ _{27} $P$ _{20} $ the heat of fusion $ \Delta H_{f} $, measured in the DSC, is 5.02 kJ g-atom$ ^{-1} $.With this value, the enthalpy difference change with undercooling between liquid and crystal is calculated as
\begin{equation}
\Delta H^{l-x}=\Delta H_{f}-\int_{T}^{T_{f}}\Delta c_{p}^{l-x}(T^{\prime})dT^{\prime},
\label{Eq12}
\end{equation}
using the measured $ \Delta H_{f} $ and $ \Delta c_{p}^{l-x} (T) $. In Fig. \ref{Fig2} the resulting enthalpy change of the Pd$ _{43} $Ni$ _{10} $Cu$ _{27} $P$ _{20} $ glass former is plotted (continuous line) in reference to that of the crystal. During undercooling experiments residual enthalpy is frozen in at the kinetic glass transition, which is represented by the dotted line.  The data represented by circles in Fig. \ref{Fig2} are experimental heats of crystallization obtained by integrating the exothermic peak during isothermal undercooling experiments of the liquid (see Ref.18 ). For example at the isothermal temperature of 693 K the $ \Delta H_{x} $ for the undercooled Pd$ _{43} $Ni$ _{10} $Cu$ _{27} $P$ _{20} $ is 3.7 kJ (g-atom)$ ^{-1} $. The Pd$ _{43} $Ni$ _{10} $Cu$ _{27} $P$ _{20} $ liquid, due to its high thermal stability in its undercooled state, is the only alloy, in which the $ c_{p} $ of the undercooled liquid and the heat of crystallization can be measured throughout the undercooled liquid simultaneously, affirming the consistency of our thermodynamic description of the undercooled liquids.

\begin{figure}[ht]
\begin{center}
\includegraphics[width=3in]{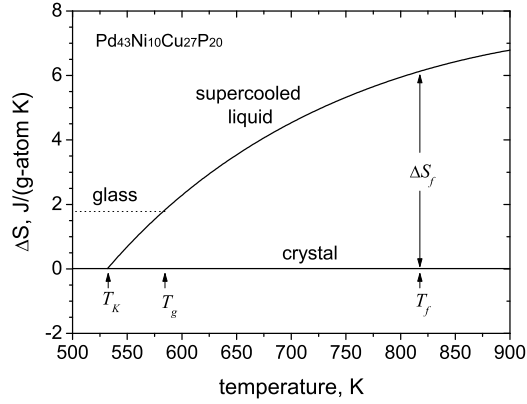}
\caption{Entropy difference of the undercooled liquid Pd$ _{43} $Ni$ _{10} $Cu$ _{27} $P$ _{20} $ with respect to the crystal as a function of temperature, calculated with Eq. (\ref{Eq13}) [18]. $ T_{K} $, $ T_{g} $, and $ T_{f} $ are the Kauzmann temperature (532 K), the calorimetric glass transition temperature for heating rates of 0.33 K/s (582 K), and the calorimetric fusion peak temperature (818 K), respectively. $ \Delta S_{f} $  is the entropy of fusion, of about 6 J g-atom$ ^{-1} $ K$ ^{-1} $ computed with Eq. (\ref{Eq14}).}
\label{Fig3}
\vspace*{-12pt}\end{center}
\end{figure}

Figure \ref{Fig3} is the plot of entropy change with undercooling for Pd$ _{43} $Ni$ _{10} $Cu$ _{27} $P$ _{20} $ glass former in reference to its crystalline counterpart, as calculated by
\begin{equation}
\Delta S^{l-x}(T)=\Delta S_{f}-\int_{T}^{T_{f}}\frac{\Delta c_{p}^{l-x}(T^{\prime})}{T^{\prime}}dT^{\prime},
\label{Eq13}
\end{equation}
where  $\Delta S_{f}$ is entropy of fusion, determined as
\begin{equation}
\Delta S_{f}=\frac{\Delta H_{f}}{T_{f}}. 	
\label{Eq14}
\end{equation}
 	 	
Note that Eq (\ref{Eq13}) and Eq (\ref{Eq6}) are not identical. $\Delta S_{f}$ is the entropy of fusion, which in BMG is not the difference in configurational entropy between crystal and liquid, as we will see later. When connecting the kinetics with the thermodynamics through the Adam-Gibbs theory [Eq. (\ref{Eq5})] it is more suitable to introduce a configurational entropy $\Delta S_{c}(T_{m}^{*})$ at some temperature $ T_{m}^{*} $ with a fixed viscosity as a fit parameter like in Eq. (\ref{Eq6}) [18].

\subsection{KINETICS}
\label{sec:vis}
\subsubsection{VISCOSITY MEASUREMENTS}

Viscosities close to the glass transition can be determined by three-point beam bending experiments. Beams with rectangular cross-sections between 0.2 and 1.0 mm$ ^{2} $ and lengths of approximately 13 mm are used as samples. A Netzsch Thermal Mechanical Analyzer (TMA 402), calibrated for heating rates of 0.025 and 0.833 K/s, according to the melting standards of indium and zinc, are used in our case to perform three-point beam-bending measurements on the samples
A beam, supported at each end by sharp edges, is subjected in the center to a constant force provided by a fused silica loading probe with a wedge-shaped head, and the corresponding deflection of the beam is measured. Using this technique viscosities ranging from $ 10^{7} $ to $ 10^{14} $ Pa s can be determined with the following equation [38]
\begin{equation}
\eta=-\frac{gL^{3}}{144I_{c}v}[M+\frac{\rho AL}{1.6}],
\label{Eq15}
\end{equation}
where $ g $ is the gravitational constant (m/s$ ^{2} $), $ I_{c} $ the cross-section moment of inertia (m$ ^{4} $), $ v $ the mid-point deflection rate (m/s), M the applied load (kg), $ \rho $ the density of the glass (kg/m$ ^{3} $), A the cross-sectional area (m$ ^{2} $) and $ L $ the support span ($ 1.196 \times 10^{-2} $ m for the apparatus). In order to increase signal-to-noise ratio, the deflection of the beam should be as large as possible, such that scatter is minimized during long-time measurements. To achieve the greatest possible deflection while remaining within the measurable rage of the apparatus, the applied load was kept constant at 0.01 kg and the cross-section of the beams varied, depending on the expected viscosity a given temperature.

\begin{figure}
\begin{center}
\includegraphics[width=3in]{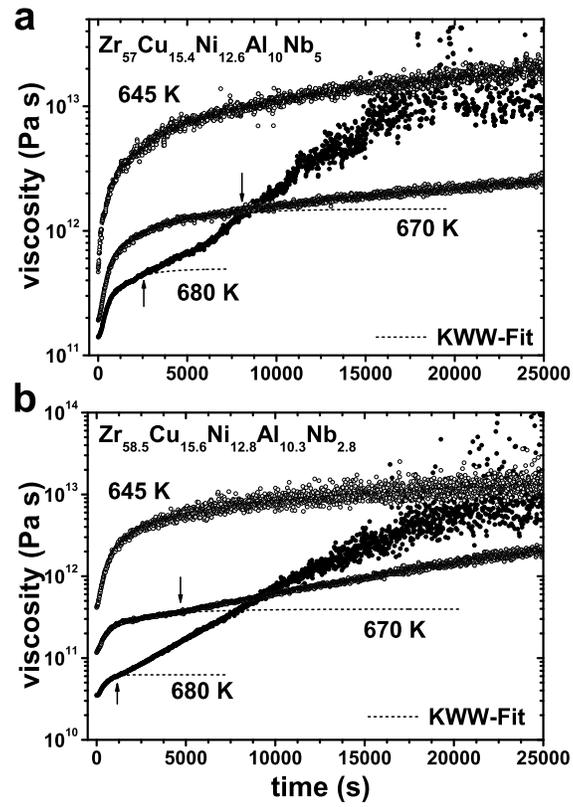}
\caption{Isothermal viscosity measurements at three different temperatures for (a) Zr$ _{57} $Cu$ _{15.4} $Ni$ _{12.6} $Al$ _{10} $Nb$ _{5} $ and (b) Zr$ _{58.5} $Cu$ _{15.6} $Ni$ _{12.8} $Al$ _{10.3} $Nb$ _{2.8} $. The relaxation from the glassy state into the equilibrium liquid is fitted the KWW equation (dashed lines). Arrows mark the onset phase separation [40].}
\label{Fig4}
\vspace*{-12pt}\end{center}
\end{figure}

\begin{sloppypar}
Figure \ref{Fig4} shows three examples of viscosities calculated from Eq. (\ref{Eq15}) at constant temperatures for Zr$ _{57} $Cu$ _{15.4} $Ni$ _{12.6} $Al$ _{10} $Nb$ _{5} $ (Vitreloy 106) and Zr$ _{58.5} $Cu$ _{15.6} $Ni$ _{12.8} $Al$ _{10.3} $Nb$ _{2.8} $ (Vitreloy 106a); these temperatures are 645 K, 670 K and 680 K. In Figs. \ref{Fig4}(a) and \ref{Fig4}(b) the viscosity at low temperatures (645 K) is seen to increase sharply from its initial value and approach a constant, equilibrium value at longer times. This is the relaxation of the glass into the equilibrium liquid region. As the temperature is increased the measured viscosity decreases, as the overall deflection rate of the sample increases due to higher atomic mobility in the glass. 
\end{sloppypar}

For both alloys, as the isothermal experiments are carried out at higher temperatures (670 K), the initial relaxation of the viscosity occurs at a shorter time and is accompanied by a gradual deviation from the equilibrium value. The arrows in Figs. \ref{Fig4}(a) and \ref{Fig4}(b) indicate the onset of deviation from equilibrium. At even higher temperatures (680 K), close to or above the calorimetric glass transition, the measured viscosity of both alloys will depart from equilibrium more rapidly and undergo a sudden increase of about two orders of magnitude from around $ 1\times 10^{11} $ Pa s to about $ 1 \times 10^{13} $ Pa s. For longer times the viscosity at this temperature remains virtually constant, suggesting that a new metastable equilibrium state has been reached. The relaxation of the sample from the glassy state into the equilibrium liquid is found to be best-described with a stretched exponential, Kohlrausch-Williams-Watts (KWW) relaxation function [39]
\begin{equation}
\eta(t)=\eta_{a}+\eta_{eq-a}\left(1-e^{-(t/\tau_{s})^{\beta}}\right), 	
\label{Eq16}				
\end{equation}
where $ \tau_{s} $ is an average shear flow relaxation time; $ \beta $, a stretching exponent; $ t $, time; and $ \eta_{a} $ the initial viscosity of the glassy alloy before relaxation. $ \eta_{eq-a} $ is the total viscosity change during relaxation from the glassy state into the equilibrium liquid. The fits of this equation to the measured viscosity data are shown as dashed lines in Figs. \ref{Fig4}(a) and \ref{Fig4}(b). The equilibrium viscosity then, $\eta_{eq}=\eta_{a}+\eta_{eq-a} $, corresponds to the constant values reached by the KWW-fits at long times.

The viscosity measurements on the amorphous samples show that, in general, these alloys exhibit a complex dependence of the viscosity on temperature and annealing time. In Fig. \ref{Fig4}(a) the measured viscosity of Zr$ _{57} $Cu$ _{15.4} $Ni$ _{12.6} $Al$ _{10} $Nb$ _{5} $ begins to depart from the fitted equilibrium value at around 7500 s for the isothermal measurement at 670 K and 2500 s for 680 K. At these temperatures the departures from equilibrium for Zr$ _{58.5} $Cu$ _{15.6} $Ni$ _{12.8} $Al$ _{10.3} $Nb$ _{2.8} $ (Fig. \ref{Fig4}b) occur earlier at around 5000 s and 1000 s, respectively. The earlier departure from the initial metastable equilibrium for Zr$ _{58.5} $Cu$ _{15.6} $Ni$ _{12.8} $Al$ _{10.3} $Nb$ _{2.8} $ can be attributed to this alloy's lower glass transition temperature compared to Zr$ _{57} $Cu$ _{15.4} $Ni$ _{12.6} $Al$ _{10} $Nb$ _{5} $ (see Ref 40). This is also explained by the relative viscosities of these two alloys at these temperatures, which for Zr$ _{57} $Cu$ _{15.4} $Ni$ _{12.6} $Al$ _{10} $Nb$ _{5} $ are around five times greater than for Zr$ _{58.5} $Cu$ _{15.6} $Ni$ _{12.8} $Al$ _{10.3} $Nb$ _{2.8} $, allowing for faster kinetics due to increased atomic mobility [40].

The equilibrium viscosities determined for these alloys are obtained by fitting Eq. (\ref{Eq16}) only to the experimental data that were taken before the onset of the deviations from equilibrium, which occur for longer times and higher temperatures. Very similar deviations from equilibrium in viscosity have been observed in the Zr$ _{41.2} $Ti$ _{13.8} $Cu$ _{12.5} $Ni$ _{10} $Be$ _{22.5} $ (Vitreloy 1) BMG [41]. In that case the deviations have been attributed to a combination of phase separation, nano-crystal formation, composition redistribution and relaxation of the remaining amorphous matrix. These examples show that one has to be very cautious interpreting viscosity measurements especially close to the glass transition due to the limited thermal stability of metallic glass formers not only with respect to crystallization but also phase separation.

\begin{figure}[h]
\begin{center}
\includegraphics[width=3in]{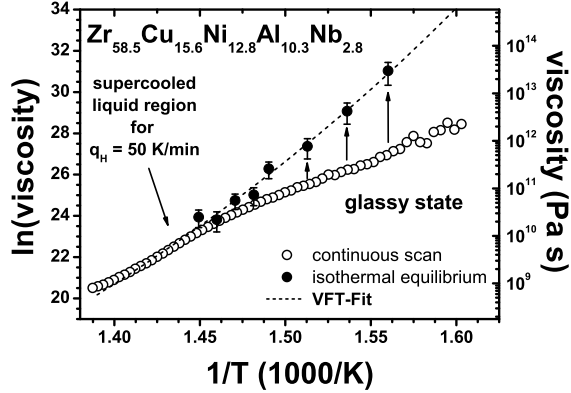}
\caption{Equilibrium viscosities from isothermal relaxation experiments (closed circles) for the Zr$ _{58.5} $Cu$ _{15.6} $Ni$ _{12.8} $Al$ _{10.3} $Nb$ _{2.8} $ as a function of inverse temperature. Also shown is the measured viscosity determined with a constant heating rate of 0.833 K/s (open circles). A single VFT fit to both sets of data in the equilibrium liquid is shown with a fragility parameter, $ D^{*} = 21.7 $ (dashed line). The arrows indicate the initial isothermal relaxation pathways that are shown in Fig. \ref{Fig4}(b) [40].}
\label{Fig5}
\vspace*{-12pt}\end{center}
\end{figure}

The temperature dependence of the equilibrium viscosity is described with the empirical VFT equation (Eq. (\ref{Eq1})). Fitting the experimental data to this equation provides the fragility parameter, $ D^{*} $, and the VFT-temperature, $ T_{0} $, the temperature at which the barriers with respect to flow would approach infinity [6]. At temperatures below the glass transition, the equilibrium liquid can be accessed through isothermal annealing of the sample. Heating with a constant rate though the glass transition and into the supercooled liquid region allows access to the equilibrium liquid at higher temperatures, before the onset of crystallization. Figure \ref{Fig5} shows a VFT-plot of the viscosity for the Zr$ _{58.5} $Cu$ _{15.6} $Ni$ _{12.8} $Al$ _{10.3} $Nb$ _{2.8} $  alloy after relaxation into the equilibrium liquid (solid circles) (see also Fig. \ref{Fig4}). The plot in Fig. \ref{Fig5} also includes the viscosity data taken from a continuous heating viscosity measurement (open circles). For the constant heating rate of 0.833 K/s the measured viscosity in the glassy state stays smaller than the equilibrium viscosity because of the frozen-in free volume – i.e. the isoconfigurational glassy state. As the sample is heated through the glass transition and into the supercooled liquid region, it leaves its isoconfigurational state and the measured viscosity corresponds to the equilibrium viscosity. A fit to the VFT equation (Eq. (\ref{Eq1})) of both the equilibrium viscosities obtained from isothermal relaxation and continuous heating well describes both sets of data in this temperature range with a fragility parameter of $ D^{*} = 21.7 $. 

The viscosity of bulk metallic glasses close to the glass transition has also been measured by parallel plate rheometry [1, 42-44]. A thin disk of metallic glass is squeezed between two circular probes. By measuring the height of the sample as a function of time the viscosity can be determined. The corresponding equation (after Stefan [45,46]) assumes only radial flow, thus being only valid in the limit of zero thickness. Therefore for each temperature several samples with different aspect ratio have to be  measured. A fit of the data to zero thickness yields the equilibrium viscosity [42]. Unfortunately, one finds a number of studies in the literature, where this care has not been taken and the viscosity has been overestimated considerably.

We determined melt viscosities in the vicinity of the equilibrium melting point of Zr-based alloys using a custom-built high-vacuum high-temperature Couette concentric cylinder viscometer, the experimental setup of which is described in Ref. [47]. The graphite shear cell is machined from Ringsdorff \textregistered-Isographite R6710. Additional microscopy investigations revealed no infiltration of the molten Zr into this specific type of graphite. Each alloy sample is first inductively heated to a temperature above $ T_{liq} $. At this temperature, a shearing profile is applied by gradually varying the shear rate, $ \dot{\gamma} $ , from $ \sim 50$ s$^{-1} $ to $\sim 450$ s$^{-1}$. The temperature is then increased in increments of 25 K, where the shearing profile is applied again. At the end of the first series of isothermal measurements, the melt is cooled back down to the initial starting temperature and the aforementioned procedure is carried out a second time.
 
The viscosity, $ \eta $, in Pa$ \cdot $s is calculated as
\begin{equation}
\eta=\frac{M}{2\pi r_{i}^{2}L\dot{\gamma}},
\label{Eq17}
\end{equation}
where $ r_{i} $ is the radius of the inner concentric cylinder in meters, $ L $ is the emersion length in meters, and $ \dot{\gamma} $ is the shear rate in s$ ^{-1} $.   The shear rate can then be calculated by 
 		
\begin{equation}
\dot{\gamma}=\frac{2\Omega r_{i}^{2}}{r_{o}^{2}-r_{i}^{2}},
\label{Eq18}
\end{equation} 							
where $ r_{i} $ and $ \Omega $ are have been defined previously and $ r_{o} $ is the outer radius of the shear cell in meters.  Equation \ref{Eq18} assumes the absence of a shear rate dependence of the viscosity and a linear shear rate profile between the inner and outer radius of the shear cell.   Due to the pronounced shear thinning behavior that we observe in Vitreloy 1, a shear rate correction factor is applied to take into consideration the non-linear shear profile that develops between the inner and outer radius of the shear cell. The corrected shear rate when measuring a non-Newtonian shear rate dependent fluid by Couette Concentric Cylinder measurements is calculated, after [48, 49], as 
 					 
\begin{equation}
\dot{\gamma}=\frac{2\Omega}{n\left(1-b^{2/n}\right)},	
\label{Eq19}			 
\end{equation}
 					 
where $ b $ is the ratio of the inner to outer radius of the shear cell and $ n $ is the exponent obtained from fitting a power law relationship to the collected torque vs. rotation data at a constant temperature.  The final shear rate deviates from the uncorrected shear rate between 0 and 40\% depending on the magnitude of the shear thinning effect.

Viscosity uncertainty was determined by calculating the statistical standard deviation of the collected data after filtering instrumental noise using a running average method.  Propagation of error through calculations is determined by applying the Kline-McClintock method [50]. 

\begin{figure}[h]
\begin{center}
\includegraphics[width=3in]{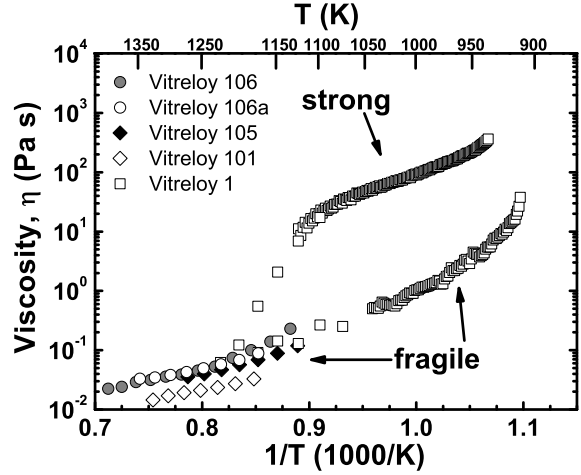}
\caption{Arrhenius-plot showing the experimentally determined, high-temperature melt viscosities of 5 Zr-based alloys. Included are the viscosity data from Vitreloy 1 ($ \Box $) taken from Ref. [47], which shows a fragile strong transition. The top horizontal axis gives the direct temperature scale for comparison [56].}
\label{Fig6}
\vspace*{-12pt}\end{center}
\end{figure}

Figure \ref{Fig6} shows viscosity data above the melting point or in the slightly undercooled liquid for 5 Zr-based alloys [56]. Included are the viscosity data from Vitreloy 1 ($ \Box $) taken from Ref. [47], which show a fragile-strong transition (see Section \ref{sec:FS}. The top horizontal axis gives the direct temperature scale for comparison. The viscosities of these equilibrium liquids of BMG are found between 0.01 and 50 Pa s depending on the fragility of the melt. This indicates, that they are considerably more fragile that silicates (about $ 10^{5} $ Pa s) but much stronger than pure metals or water, which have melt viscosities of about $ 3\times 10^{-3} $ Pa s.

A second method to measure the viscosity of the liquid close to the melting point is performed in an Electrostatic Levitator (ESL) [51-53]. The levitated droplet is brought to oscillations. From the damping the viscosity is determined [54,55]. This method has the advantage that it is in a clean environment and slip of the melt or contamination in the rotating cups is not an issue. However the shear rate is determined by the size of the droplet and thus can be adjusted only slightly by changing the size of the droplet.

\subsubsection{HEATING RATE DEPENDENCE OF THE GLASS TRANSITION}

\begin{figure}
\begin{center}
\includegraphics[width=3in]{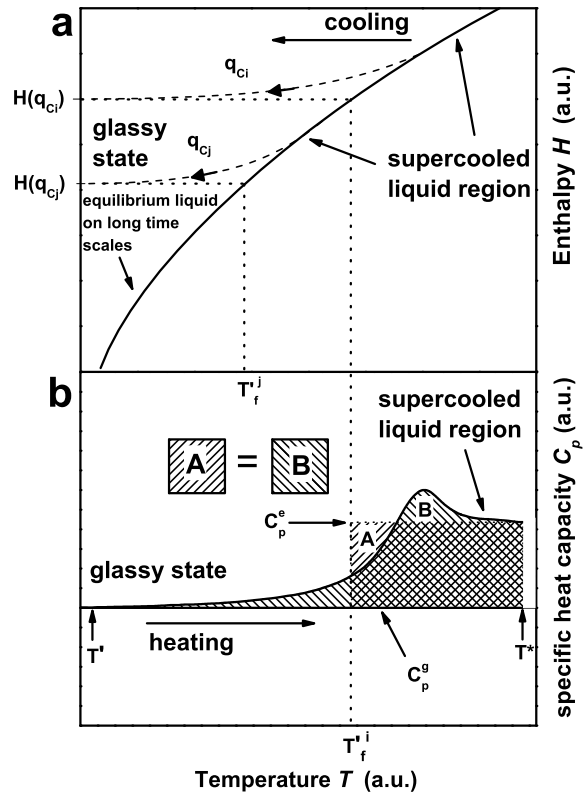}
\caption{(a) Schematic plot showing enthalpy, $ H $, versus temperature, $ T $, during the formation of different glassy states, $ H(q_{Ci}) $ and $ H(q_{Cj}) $ by undercooling with rates $ q_{Ci} $ and $ q_{Cj} $, respectively, where $ q_{Ci}>q_{Cj}$. The corresponding limiting fictive temperatures, $ T^{\prime i}_{f} $ and $ T^{\prime j}_{f} $ are shown as projections (dotted lines) of the glass curve (dashed lines) onto the equilibrium liquid line (solid line) such that $ T^{\prime i}_{f} > T^{\prime j}_{f} $. (b) Schematic of the specific heat capacity, $ C_{p} $, versus temperature, $ T $, during heating of the glass that was previously cooled from the liquid state with a rate $ q_{Ci} $. The determination of $ T^{\prime i}_{f} $ is also shown here [63]. }
\label{Fig7}
\vspace*{-12pt}\end{center}
\end{figure}

Upon undercooling from the liquid, a unique structural configuration is frozen into the glassy state as the liquid falls out of equilibrium at a certain temperature. This temperature is known as the glass transition temperature, $ T_{g} $, and is a unique function of the cooling rate, $ q_{C} $ [57]. The only unambiguous definitions of the glass transition temperature are those that are determined during cooling and depend only on the cooling rate [58]. The concept of a characteristic glass transition temperature as being a unique function of the cooling rate was proposed by A. Q. Tool in 1946 as the fictive glass transition temperature [59]. A distinct temperature is defined, on cooling, that is directly associated with the limiting value of the quantity measured to fall out of equilibrium at the glass transition. This temperature is known as the limiting fictive temperature, or $ T^{\prime}_{f} $, and is defined as the glass transition temperature as measured on cooling [57].  Figure \ref{Fig7}(a) shows a schematic representation of the enthalpy, $ H $, during the formation of a glass during undercooling. Geometrically, $ T^{\prime}_{f} $ is defined from a point well into the glassy region. It is the temperature of intersection on the equilibrium $ H-T $ curve with a line drawn through the point of interest inside the glassy state having a slope equal to that of the glass curve. $ T^{\prime}_{f} $ is usually determined from a DSC up-scan using the definition put forth by Moynihan [57]:

\begin{equation}
\int_{T^{*}}^{T_{f}^{\prime}}(C_{p}^{e}-C_{p}^{g})dT_{f}=\int_{T^{*}}^{T^{\prime}}(C_{p}-C_{p}^{g})dT.
\label{Eq20}
\end{equation}

The curves $ C_{p}^{e} $ and $ C_{p}^{g} $ represent the heat capacities belonging to the equilibrium liquid and glassy states, respectively. $ T^{*} $ is any temperature above the glass transition where $ C_{p} = C_{p}^{e} $, and $ T^{\prime} $ is a temperature well below the glass transition and into the glassy state where $ C_{p} = C_{p}^{g} $. This construction is shown schematically in Fig. \ref{Fig7}(b). $ T^{\prime i}_{f} $ is determined here graphically by matching the area underneath the curve with that of a rectangle defined by $ C_{p}^{e} $ and $ C_{p}^{g} $. The heating rate, $ q_{H} $, is intentionally left ambiguous, as $ T^{\prime i}_{f} $ depends only on the cooling rate, $ q_{Ci} $.

In calorimetric experiments, the DSC up-scan is used to determine the glass transition temperature, which is defined during heating of the sample with a heating rate, $ q_{H} $. The shift of the glass transition temperature with the heating rate, $ q_{H} $, is assumed to reflect the fragility of the material [57, 60, 61], and the fragility parameter has been determined by fitting the heating rate dependence of the glass transition with Eq. (\ref{Eq1}) [1, 3, 62, 63].

\begin{figure}[h]
\begin{center}
\includegraphics[width=3in]{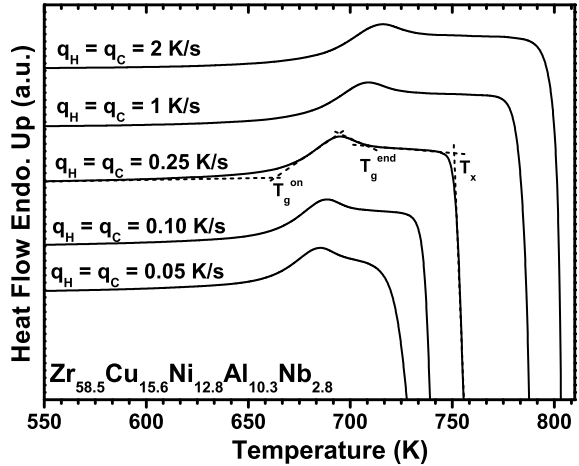}
\caption{DSC heat flow curves of the Zr$ _{58.5} $Cu$ _{15.6} $Ni$ _{12.8} $Al$ _{10.3} $Nb$ _{2.8} $ alloy showing the glass transition shift for the same heating and cooling rates; $ q_{H} $ and $ q_{C} $, respectively. Included are the definitions of the onset and end of the glass transition, as well as the onset of the crystallization event – $ T_{g}^{onset} $, $ T_{g}^{end} $ and $ T_{x}$, respectively. Curves were translated along the vertical axis for easier comparison [63]. }
\label{Fig8}
\vspace*{-12pt}\end{center}
\end{figure}

During heating from the glassy state and into the supercooled liquid region, one can define a particular average structural relaxation time, $\tau$, for each heating rate, $ q_{H} $, such that

\begin{equation}
\tau=\frac{\Delta T}{q_{H}},
\label{Eq21}
\end{equation}
where $ \Delta T = T_{g}^{end}-T_{g}^{onset} $ (Fig. \ref{Fig8}) is the width of the glass transition. Each value of $\tau$ is inversely proportional to $ q_{H} $ and a plot of $ T_{g} $ versus $ \tau $ will show the kinetic shift of the glass transition, where lower glass transition temperatures are measured for slower heating rates [1]. The kinetic fragility parameter $ D^{*}_{\tau} $ can then be determined from a fit of this shift to Eq. (\ref{Eq1}).

The convention of using the same heating rate, $ q_{H} $, as that of an immediately preceding cooling rate, $ q_{C} $, from the supercooled liquid region, has been adopted by many investigators [60, 64, 65] but is still not considered standard amongst all researchers in the field of BMG. As we showed in Ref. [63], the glass transition temperature as measured on heating can vary greatly depending on whether or not the convention $ q_{H} = q_{C} $ is kept. Furthermore, if $ q_{H} \neq q_{C} $, this will lead to different apparent values of $ D^{*} $ being obtained for the same glass-former. The onset temperature of the glass transition, as measured on heating, is not only dependent on the heating rate, but is also sensitive to the initial structural state "frozen-in" to the glass during cooling from the liquid [58]. As such, it is important to access the effect of the material's structural state on the measurement of the glass transition during heating. Furthermore, since the shift of the onset temperature of the glass transition with the heating rate reflects the fragility of the material, an accurate measurement of this temperature is necessary in order to determine the correct fragility.

\subsubsection{RELAXATION CLOSE TO THE GLASS TRANSITION}

If bulk metallic glasses are heat treated below the glass transition, they structurally relax into a state that corresponds to the supercooled liquid when observed on a long time scale. This can be observed as a change in viscosity (see above Section \ref{sec:vis}), volume, enthalpy and even leads to an apparent $ c_{p} $ for the supercooled liquid on a long time scale.

Besides the already described isothermal viscosity measurement, enthalpy relaxation can be carried out in the DSC by first heating each sample (with masses ranging from 80 to 100 mg) to the desired annealing temperature, before the onset of the calorimetric glass transition, with a rate of e.g., 0.416 K s$ ^{-1} $, and then holding isothermally for a certain amount of time. At each annealing temperature the samples are held for various times. The maximum annealing time at each temperature should be chosen to be long enough to completely relax the samples (see, e.g. Ref. [33]) and ensure relaxation into the equilibrium liquid while avoiding the crystallization events measured isothermally in Ref. [66]. After completion of the anneal the samples are first cooled to room temperature with a rate of 0.416 K s$ ^{-1} $ and then subsequently heated with the same rate throughout the glass transition, where the enthalpy recovery is measured, and past the crystallization event to a temperature of 853 K.

\begin{figure}[h]
\begin{center}
\includegraphics[width=3in]{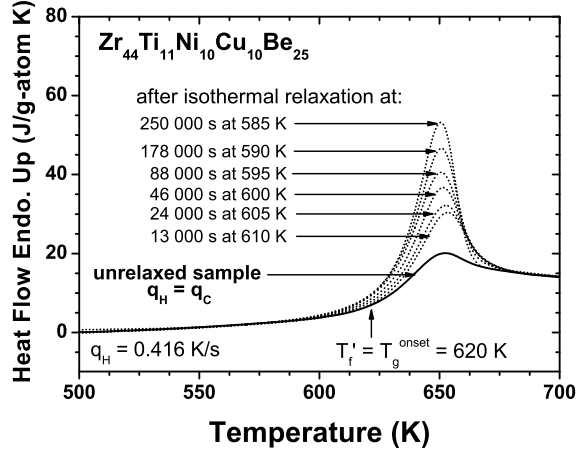}
\caption{Enthalpy recovery curves (dotted lines) after isothermal relaxation into the equilibrium liquid at the specified temperatures. The solid line represents the curve of an unrelaxed sample, i.e. heated with the same rate, $ q_{H} $, as that of an immediately preceding cooling, $ q_{C} $, from the supercooled liquid region. $ T_{f}^{\prime} $ is the limiting fictive temperature and approximates $ T_{g}^{onset} $ when $ q_{H} = q_{C} $ [67].}
\label{Fig9}
\vspace*{-12pt}\end{center}
\end{figure}

\begin{figure}[h]
\begin{center}
\includegraphics[width=3in]{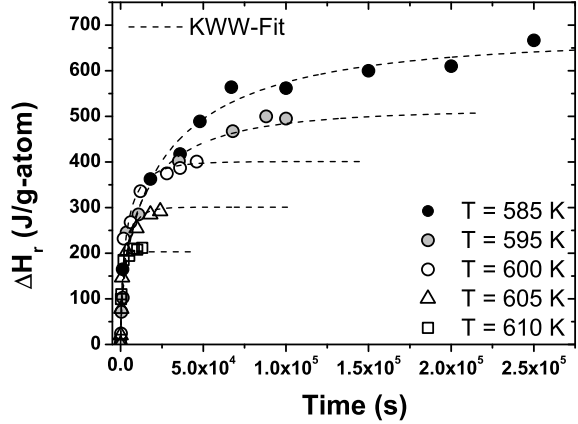}
\caption{Experimentally determined enthalpies of recovery, $ \Delta H_{r} $, after isothermal annealing of the Zr$ _{44} $Ti$ _{11} $Ni$ _{10} $Cu$ _{10} $Be$ _{25} $ BMG at various times for the temperatures shown. The dashed lines represent the fits of the KWW-equation to the experimental data.  The error is on the order of the symbol size [67].}
\label{Fig10}
\vspace*{-12pt}\end{center}
\end{figure}

In Fig. \ref{Fig9} the enthalpy recovery curves of the completely annealed samples (dotted lines) are shown for the Zr$ _{44} $Ti$ _{11} $Ni$ _{10} $Cu$ _{10} $Be$ _{25} $ BMG after heating with the rate $ q_{H} = 0.416 $ K s$ ^{-1} $ throughout the glass transition [67]. The recovery curves are shown alongside a scan of the unrelaxed sample (solid line), i.e., a sample that was heated with the same rate,  $ q_{H}$, as that of an immediately preceding cooling, $ q_{C}$, from the supercooled liquid region. It was established previously that if the convention $ q_{H} =  q_{C}$ is held, the measured onset temperature of the glass transition on heating, $ T_{g}^{onset} $, approximates the limiting fictive temperature [63]. The amount of enthalpy recovered, $ \Delta H_{r} $, after heating throughout the glass transition is calculated as the area between the respective recovery curve and that of the unrelaxed sample at a heating rate of $ q_{H} = 0.416 $ K s$ ^{-1} $:

\begin{equation}
\Delta H_{r}=\int_{500 K}^{700 K}\left[\left(\frac{dQ}{dt}\right)_{a}-\left(\frac{dQ}{dt}\right)_{u}\right]AdT,
\label{Eq22}
\end{equation}
where $ (dQ/dt) _{a}$ and $ (dQ/dt)_{u} $ are the DSC heat flow signals of the annealed and unrelaxed samples, respectively, in units of mW. Using the constant $ A=\mu m^{-1} q_{H}^{-1} $, where $ \mu $ is the gram-atomic mass of the sample and $ m $ is the sample's mass in mg, the value of $ \Delta H_{r} $ is determined in units of J g-atom$ ^{-1} $. At 700 K all samples are equilibrated in the metastable, supercooled liquid region. In Fig. \ref{Fig10} the values of $ \Delta H_{r} $ calculated using Eq. (\ref{Eq22}) are shown for various annealing times at the selected temperatures. For a given annealing temperature, the enthalpy recovery as a function of time, $ \Delta H_{r} (t)$, approaches a constant value as the sample relaxes into equilibrium at longer annealing times. As described in paragraph 2.2.1 relaxation processes in amorphous materials are usually found to be best described with a KWW stretched exponential function [68-71] of the general form

\begin{equation}
\phi(t)=\phi_{eq} (1-e^{-(t/\tau)^{\beta_{KWW}}}),
\label{Eq23}
\end{equation}
where $ \phi(t) $ is the relaxing quantity and $ \phi_{eq} $ is the value of the relaxing quantity at equilibrium in the supercooled liquid as $ t \rightarrow \infty$. Here, $ t $ is the time, $\tau$ a characteristic relaxation time and $ \beta_{KWW} $ is the stretching exponent parameter ($ 0 < \beta_{KWW} < 1 $). The fitting of the experimental data to the function in Eq. (\ref{Eq23}) was carried out using a Chi-squared minimization algorithm and the results are shown by the dashed lines in Fig. \ref{Fig10}.

Volumetric measurements of the relaxation below $ T_{g} $ were carried out in the TMA (dilatometer mode) using a vertical, fused silica loading probe. Rectangular samples with dimensions of approximately $ 2 \times 2 \times 8 $ mm were cut from the rods and used for the dilatometric measurements. The glassy samples were heated with a rate of 0.416 K s$ ^{-1} $ to the desired temperatures and then held isothermally where the length relaxation was directly measured. The load on the sample's surface was supplied by a spring-loaded linear variable differential transformer and was calculated to be $ 0.20  \pm 0.04 $ mN. 

\begin{figure}[h]
\begin{center}
\includegraphics[width=3in]{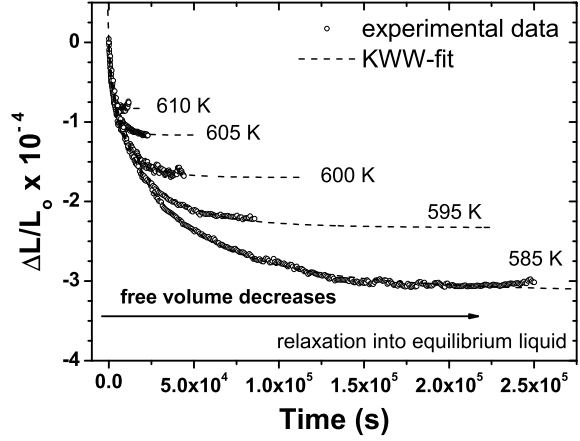}
\caption{Experimental relative changes in length, $ \Delta L/L_{o} $, of amorphous samples of Zr$ _{44} $Ti$ _{11} $Ni$ _{10} $Cu$ _{10} $Be$ _{25} $ during relaxation into the equilibrium liquid for the annealing temperatures shown (open circles). The fits of the experimental data to the stretched exponential (KWW) function are also shown (dashed lines) [67].}
\label{Fig11}
\vspace*{-12pt}\end{center}
\end{figure}

Figure \ref{Fig11} shows the relative change in length, $ \Delta L/L_{o}(t) $, of the amorphous samples as they are relaxed from the glassy state into the equilibrium liquid region during isothermal annealing at the temperatures indicated. It can be seen that, at lower temperatures, the relative changes in length are greater than at higher temperatures closer to the glass transition. The experimental data in Fig. \ref{Fig11} (open circles) are fitted with a KWW function of the form in Eq. (\ref{Eq23}).

If no temperature changes occur during the relaxation, conventional thermal expansion effects can be discounted and the measured reduction in volume is attributed solely to the reduction in excess free volume of the glass. Furthermore, assuming that structural relaxation occurs isotropically, the relative change in free volume of the amorphous sample, $ \Delta v_{f} /v_{m} $, is given by its relative change in length, $ \Delta L/L_{o}(t) $ [72-74]:

\begin{equation}
\frac{\Delta v_{f}}{v_{m}}=3\frac{\Delta L}{L_{o}}.
\label{Eq24}
\end{equation}

\begin{figure}[h]
\begin{center}
\includegraphics[width=3in]{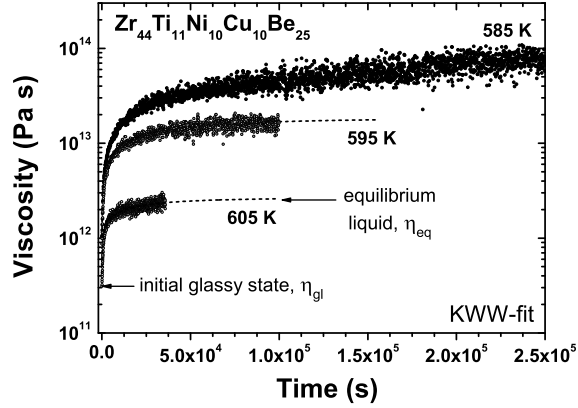}
\caption{Isothermal viscosity measurements on Zr$ _{44} $Ti$ _{11} $Ni$ _{10} $Cu$ _{10} $Be$ _{25} $ at three selected temperatures below $ T_{g} $ (585, 595 and 605 K). The relaxation from the initial glassy state into the equilibrium liquid is fitted with the stretched exponential (KWW) equation (dashed lines) [67].}
\label{Fig12}
\vspace*{-12pt}\end{center}
\end{figure}

The equilibrium viscosities of the Zr$ _{44} $Ti$ _{11} $Ni$ _{10} $Cu$ _{10} $Be$ _{25} $ alloy were determined in independent measurements in the vicinity of the glass transition using the three-point beam-bending method described in Sec. 2.2.1. Figure \ref{Fig12} shows the experimental isothermal data for three selected temperatures: 585 K, 595 K and 605 K. All samples were heated to their respective annealing temperatures with a rate of 0.416 K s$ ^{-1} $ and held there until equilibrium was reached. Fits of Eq. (\ref{Eq23}) to the measured viscosity data at selected temperatures are shown as dashed lines in Fig. \ref{Fig12}. In this equation $ \phi_{eq} $ is taken to be the equilibrium viscosity, $ \eta_{eq}=\eta_{gl}+\Delta \eta $, where $ \eta _{gl} $ is the initial viscosity of the glassy alloy before relaxation and $\Delta \eta $ is the viscosity increase during relaxation from the glassy state into the equilibrium liquid. $ \eta_{eq}$ therefore corresponds to the constant value reached by the KWW-fits at long times.

\begin{figure}[h]
\begin{center}
\includegraphics[width=3in]{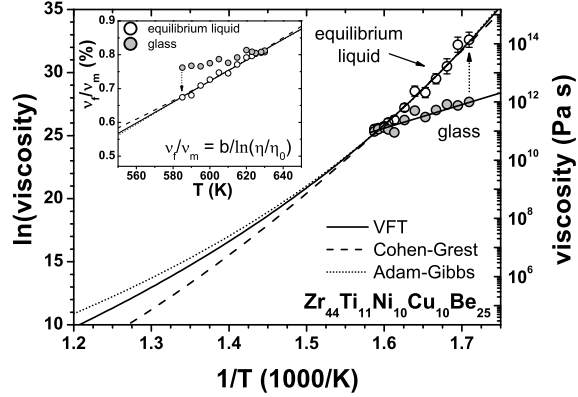}
\caption{Equilibrium viscosities (open circles), as well as the viscosities of the glass immediately prior to relaxation (shaded circles) as a function of inverse temperature for the Zr$ _{44} $Ti$ _{11} $Ni$ _{10} $Cu$ _{10} $Be$ _{25} $ alloy. The fits to the experimental data using Eqs. (\ref{Eq1}), (\ref{Eq2}) and (\ref{Eq5}) are shown as the solid, dashed and dotted lines, incorporating the expression for the relative free volume given in Eqs. (\ref{Eq3}), (\ref{Eq4}) and (\ref{Eq7}), respectively. Using Eq. (\ref{Eq2}) (inset), the relative free volumes are determined from the experimental viscosity data for the equilibrium and glassy states (see inset; open and shaded circles, respectively). The curves shown in the inset are the relative free volumes calculated from the fits shown in the main figure. The dotted arrows schematically show the path of relaxation into a more viscous amorphous state with lower free volume [67].}
\label{Fig13}
\vspace*{-12pt}\end{center}
\end{figure}

The equilibrium viscosity data (open circles) are shown in Fig. \ref{Fig13} along with the viscosities of the glassy alloy immediately before relaxation (shaded circles). A non-linear fit of the VFT-equation (Eq. (\ref{Eq1})) was performed to the equilibrium data (solid line), giving the fragility parameter, $ D^{*} = 25.4 $ and the VFT-temperature, $ T_{0} = 366.6 $ K. Additionally, a fit of the Doolittle equation (Eq. (\ref{Eq2})) incorporating the expression for the free volume according to Cohen and Grest (Eq. (\ref{Eq4})) was performed (dashed line), giving the fit parameters $ bv_{m}\varsigma_{0}/k = 5000.6 $ K, $ T_{q} = 666.6 $ K and $ 4v_{a}\varsigma_{0}/k = 160.7 $ K. Finally, a fit of the equilibrium viscosity data to the Adam-Gibbs equation (Eq. (\ref{Eq5})) is also shown (dotted line), resulting in the fit parameters $ C = 320.17 $ kJ g-atom$ ^{-1} $ in Eq. (\ref{Eq5}) and $ S_{c}(T_{m}^{*}) = 18.27 $ J g-atom$ ^{-1} $ K$ ^{-1} $ in Eq. (\ref{Eq6}).

The relative free volume, $ v_{f}/v_{m} $, as a function of temperature was calculated using Eqs. (\ref{Eq3}), (\ref{Eq4}) and (\ref{Eq7}) and is shown from 550 to 650 K in the inset in Fig. \ref{Fig13}. The relative free volumes of the equilibrium liquid and glassy states were calculated from the experimental viscosity data using Eq. (\ref{Eq2}) and are shown in the inset (open and shaded circles, respectively). The Doolittle parameter, $ b $, was determined to be 0.288, using the relation $ \alpha _{f} = b/(D^{*}T_{0}) $, where $\alpha_{glass}$ was measured here as $ 2.22 \times 10^{-5} $ K$ ^{-1} $ using a dilatometric method and the value of $ \alpha_{liq} $ was taken to be the same as that for Vitreloy 1$ ^{TM} $ from Ref. [75].

\begin{figure}[h]
\begin{center}
\includegraphics[width=3in]{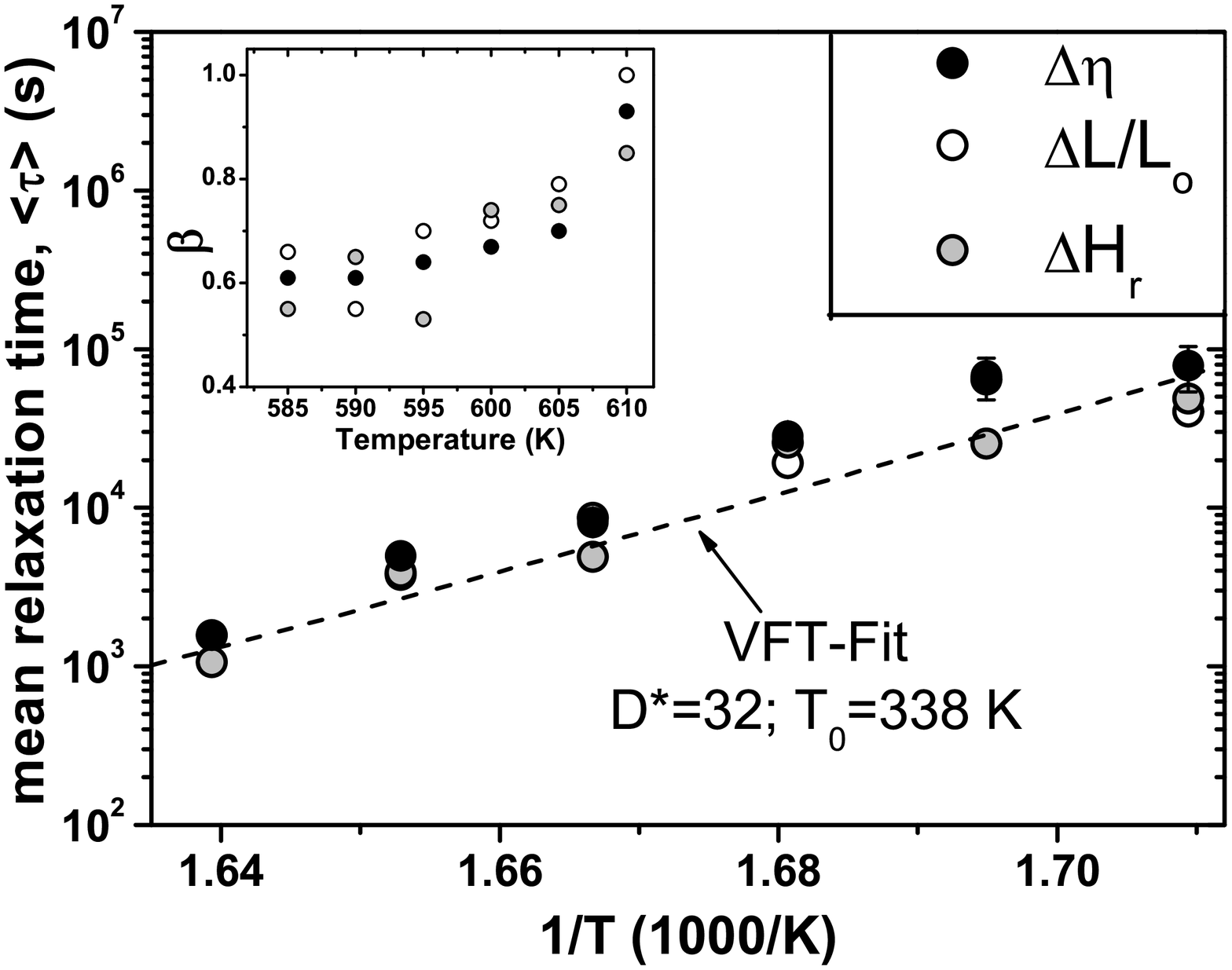}
\caption{Mean characteristic relaxation times, $ <\tau> $, obtained from fitting Eq. (\ref{Eq23}) to the experimental data taken at various annealing temperatures below the glass transition. Shown are the relaxation times taken from fitting the change in viscosity, $\Delta \eta$ (filled circles), change in relative length, $ \Delta L/L_{o} $ (open circles) and enthalpy recovery, $ \Delta H_{r} $ (shaded circles). The error is on the order of the symbol size, unless otherwise given. A fit of the VFT-equation (dashed line) to the experimental data is also shown, corresponding to the parameters $ D^{*} $ = 32 and $ T_{0} $ = 338 K. The stretching exponent parameter, $\beta _{KWW}$, reaches unity in the proximity of the glass transition (inset) [67].}
\label{Fig14}
\vspace*{-12pt}\end{center}
\end{figure}

The data shown in Fig. \ref{Fig14} correspond the characteristic relaxation times, $ \tau $, and $ \beta_{KWW} $-values (inset) obtained from fitting Eq. (\ref{Eq23}) to the experimental data in Figs. \ref{Fig10}-\ref{Fig12}). In Fig. \ref{Fig14} there is very good agreement between the values of $ \tau $ determined at each annealing temperature for each set of data: the change in viscosity, $ \Delta \eta $ (filled circles), change in relative length, $ \Delta L/L_{o} $ (open circles) and enthalpy recovery, $ \Delta H_{r} $ (shaded circles). This gives a direct link between each of the relaxing quantities and shows that the volumetric changes observed here during structural relaxation (Fig. \ref{Fig11}) can be attributed to the changes in free volume (Figs. (\ref{Fig12}) and (\ref{Fig13})).

\section{KINETIC FRAGILITY, THERMODYNAMICS AND ADAM GIBBS THEORY}

\begin{figure}[h]
\begin{center}
\includegraphics[width=3in]{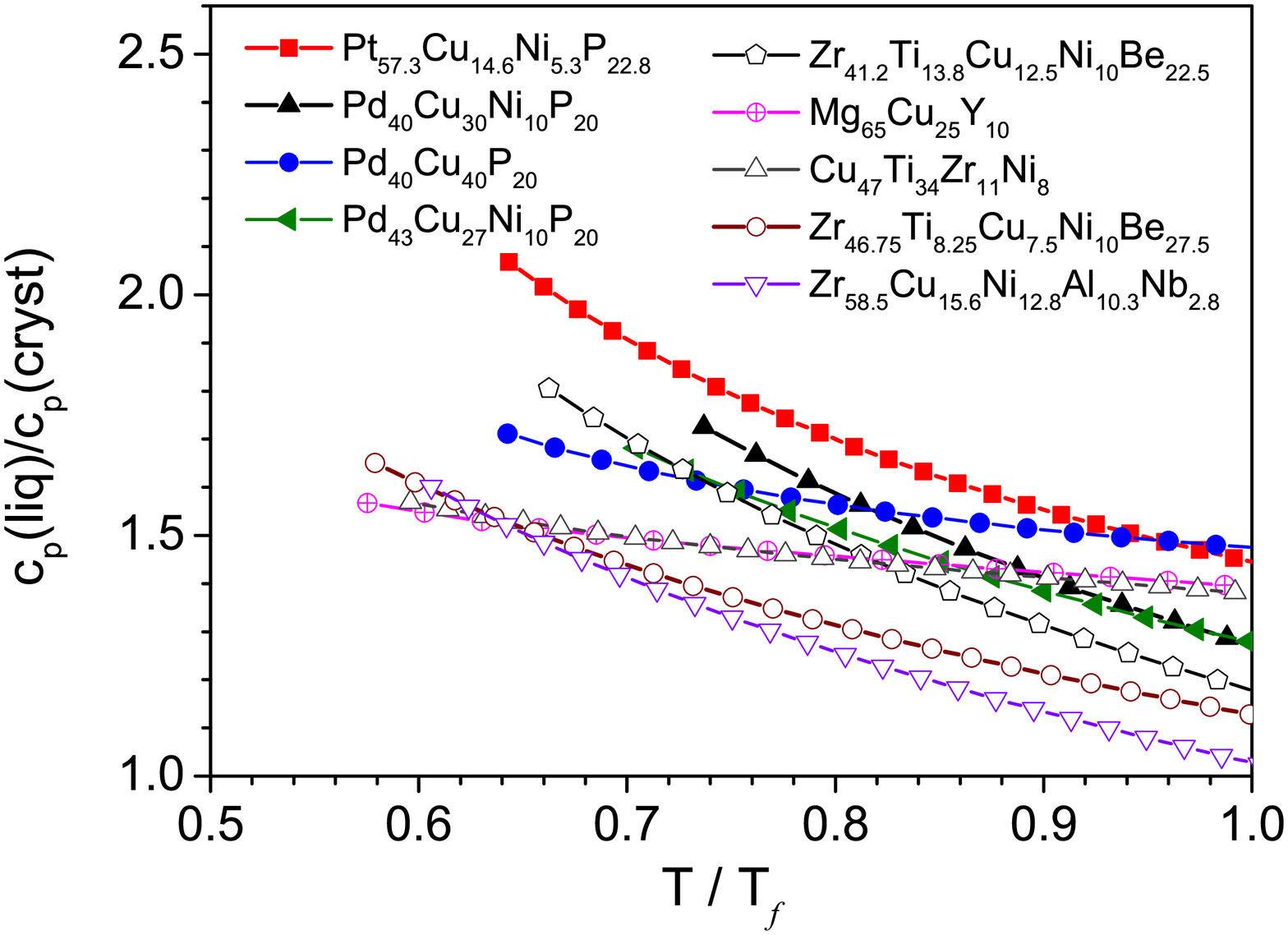}
\caption{Plot of specific heat capacity ratio between the liquid and the crystal for the selected alloys, as calculated from Eqs. (\ref{Eq6}), as a function of temperature normalized to the calorimetric fusion peak temperature $ T_{f} $ [18].}
\label{Fig15}
\vspace*{-12pt}\end{center}
\end{figure}

The thermodynamic data of 9 glass formers have been extracted from the literature and have been analyzed in a similar way as above [18]. In Fig. \ref{Fig15}, the specific heat capacity ratio  $ c_{p}(liq)/c_{p}(cryst) $ of each of the 6 glass formers of this study is plotted as a function of $ T_{f} $ - normalized temperature. The functions are calculated with Eqs. (\ref{Eq10}) and (\ref{Eq11}) after fitting the raw specific heat capacity data as described in the paragraph 2.1.  The values of $ T_{f} $ and fitting parameters $ a $, $ b $, $ c $, $ d $ can be found in Ref. 18.

\begin{figure}[h]
\begin{center}
\includegraphics[width=3in]{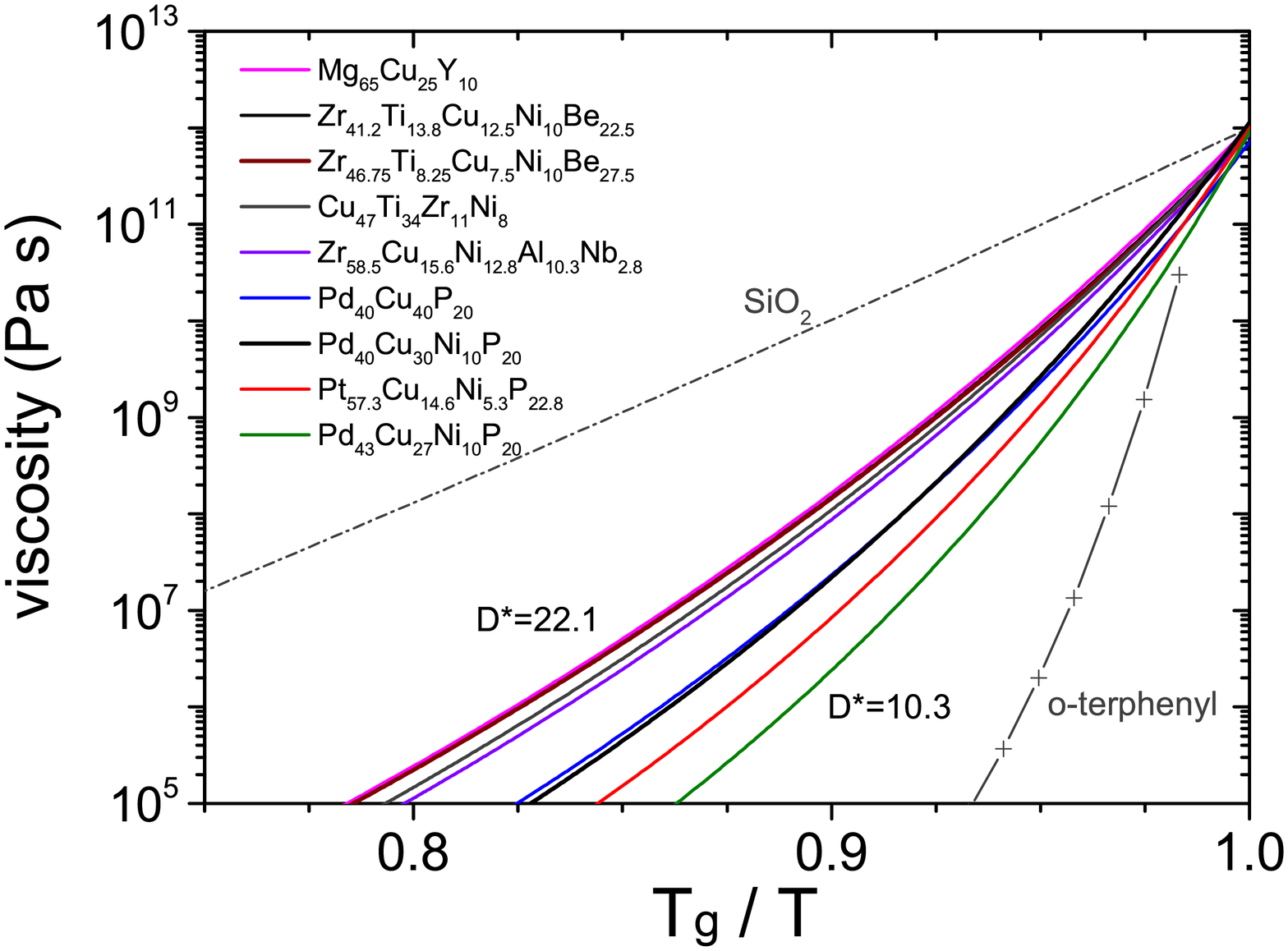}
\caption{Fragility plot of all of the selected bulk metallic glass forming liquids. The continuous lines are the fits to VFT- equation [Eq. (\ref{Eq1})]. $ D^{*} $ is the fragility parameter [18].}
\label{Fig16}
\vspace*{-12pt}\end{center}
\end{figure}

Figure \ref{Fig16} shows the low temperature range of $ T_{g} $-scaled VFT plot that includes all of the glass formers of this study. The curves are VFT fits with $ D^{*} $ and $ T_{0} $ parameters listed in Ref 18. In this study only low temperature data, around $ T_{g} $, are considered which were determined by isothermal three-points beam bending (ITPBB), parallel plate rheometry (PPR), and isothermal creep rheometry (ICR). Data at high temperatures (above the liquidus) such as those deriving from electrostatic levitation or high-temperature contact rheometry were not considered. According to Way at al. [47] glass formers may be fragile in the molten state and undergo a fragile to strong transition during undercooling. This has been directly measured by rotating concentric-cup rheometry on Zr$ _{41.2} $Ti$ _{13.8} $Cu$ _{12.5} $Ni$ _{10.0} $Be$ _{22.5} $. A recent investigations on other Zr-based systems as well as Fe-based systems show similar behaviors [56] and suggest that this may be a common phenomenon among BMG formers. This will be discussed below in more detail.  

\begin{figure}[h]
\begin{center}
\includegraphics[width=3in]{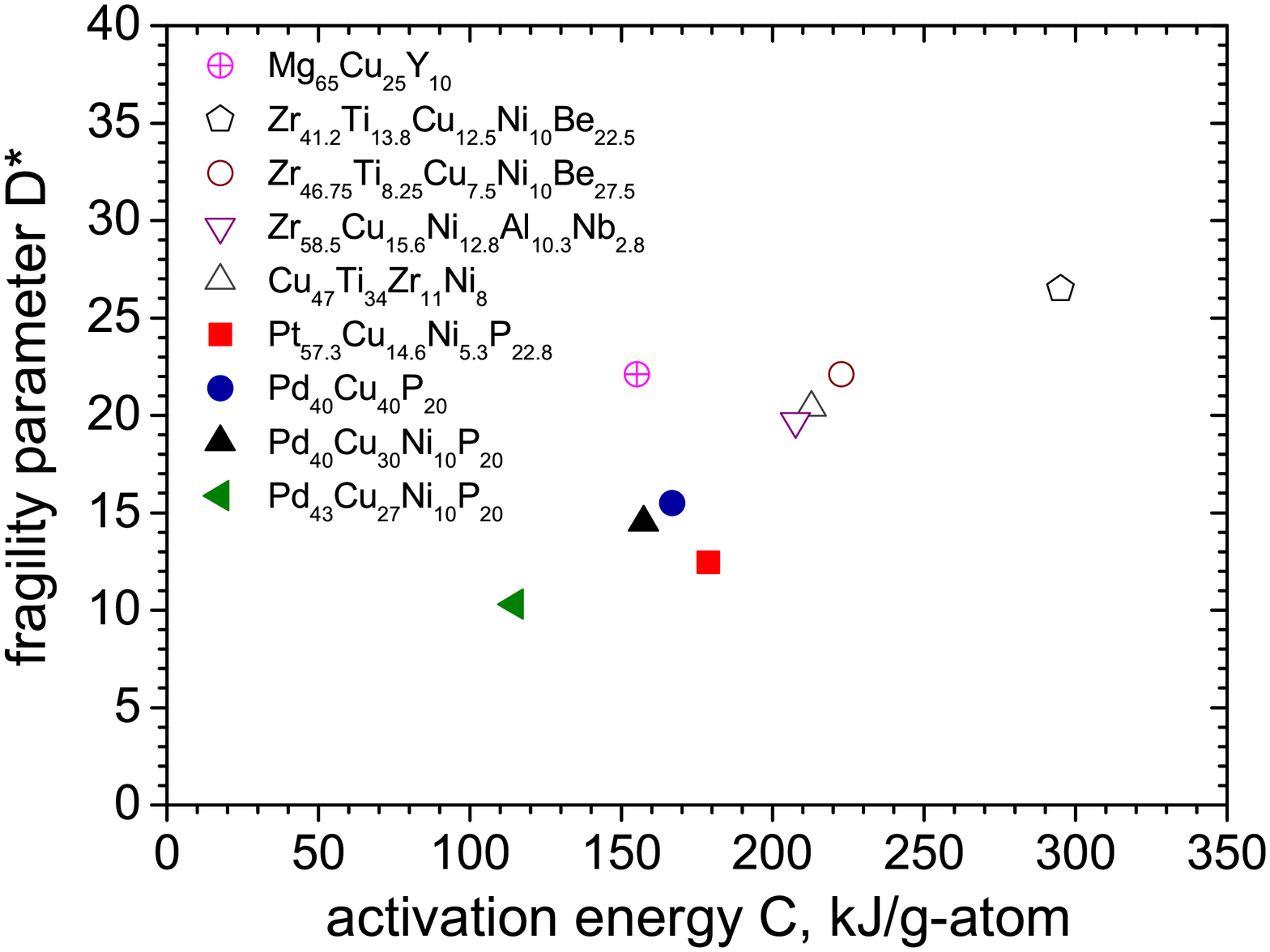}
\caption{Plot of the fragility parameter $ D^{*} $ against free activation energy per particle to cooperative rearrangements $ C $ for the selected bulk metallic glass forming liquids [18].}
\label{Fig17}
\vspace*{-12pt}\end{center}
\end{figure}

Identical data sets used for VFT fitting are fitted to the Adam-Gibbs equation. The fitting parameters $ C $ and $ S_{c}(T_{m}^{*}) $ are optimized until Eq. (\ref{Eq5}) shows the same rise in viscosity of the VFT fits in a plot of viscosity vs. ($ 1/T $). The obtained thermodynamic parameters are listed in Ref.18. The free enthalpy barrier per particle to cooperative rearrangements, $ C $, ranges between 150 and 300 kJ g-atom$ ^{-1} $ K$ ^{-1} $ which compares well with the values of activation enthalpy for diffusion measured by Faupel [75]. $ C $ progressively increases with decreasing fragility. The stronger the glass, the larger is the $ C $ as shown in Fig. \ref{Fig17}. In this plot the fragility parameter $ D^{*} $ increases monotonically with increasing activation energy $ C $. 

\begin{figure}[h]
\begin{center}
\includegraphics[width=3in]{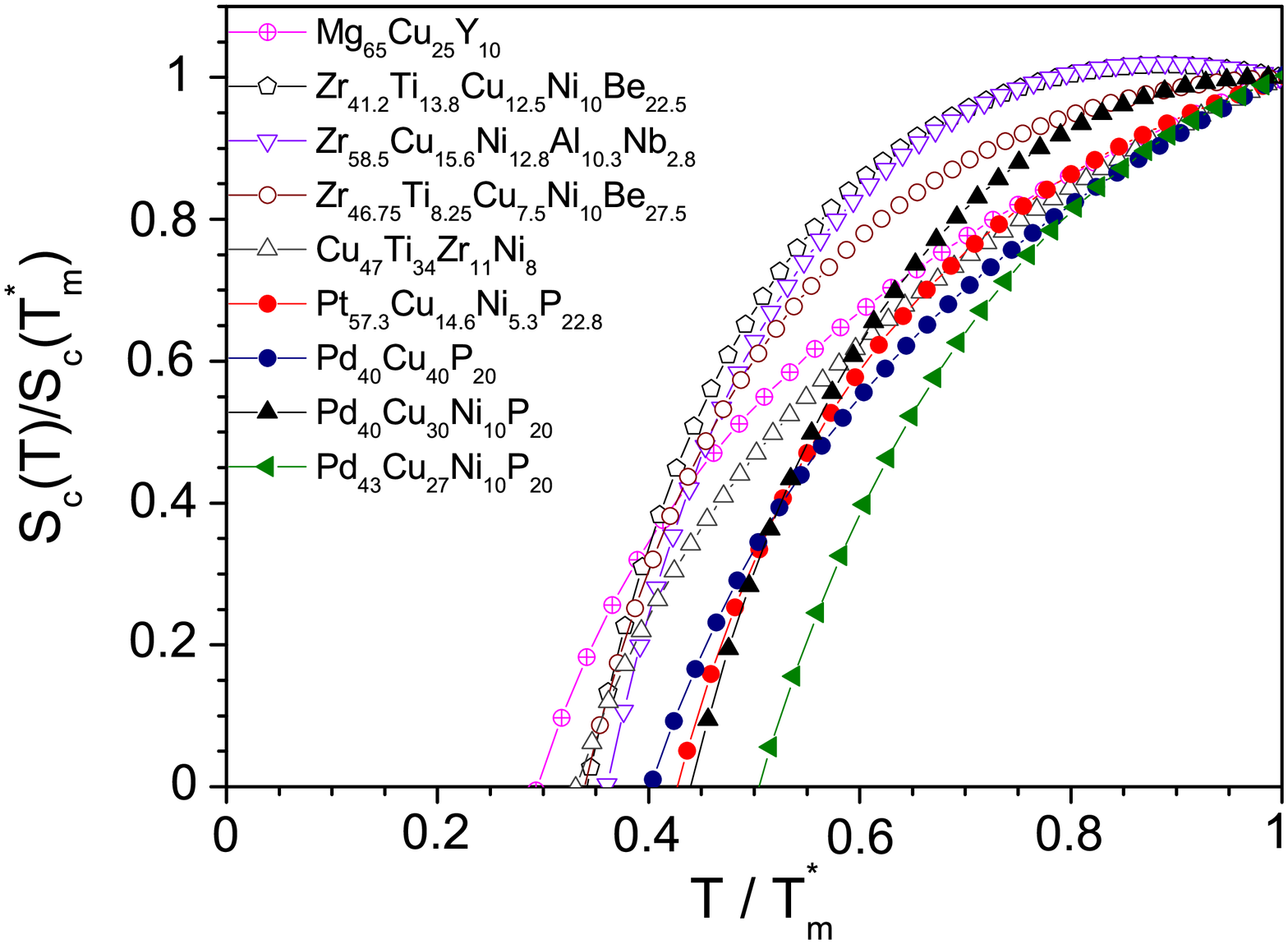}
\caption{Normalized configuration entropy plot of the selected bulk metallic glass forming liquids as a function of temperature. The normalization is against the configurational entropy of the liquid for a temperature, $ T_{m}^{*} $, at which the viscosity is of 1 Pa s. The temperature at which the configurational entropy of the liquid, Sc, vanishes is about $ T_{m}^{*}/3 $ of for all of the studied Zr-base BMG alloys, whereas for the Pt- and the Pd- base BMGs the Sc vanished at higher temperature, between $ T_{m}^{*}/2.5 $ and $ T_{m}^{*}/2 $ [18].}
\label{Fig18}
\vspace*{-12pt}\end{center}
\end{figure}

The configurational entropy of each glass former is calculated with Eq. (\ref{Eq6}) and plotted normalized by $ S_{c}(T_{m}^{*}) $ in Fig. \ref{Fig18} as a function of $ T_{m}^{*} $-scaled temperature. The temperature at which the configuration entropy of the liquid vanishes is determined by setting Eq. (\ref{Eq6}) equal to zero. For each glass former this temperature, denoted as $ T_{0}^{*} $, is represented by the intercepts with the x-axis in Fig. \ref{Fig18}. The $ T_{0}^{*} $ values for each glass former of this study are listed in Ref.18. 

The fragility plot of Fig. \ref{Fig16}, shows that the Zr- containing alloys are the strongest ($ D^{*}\sim 20-22 $), comparable to  Mg$_{65}$Cu$_{25}$Y$_{10}$ ($ D^{*}=22.1 $). The Pt-based and the Pd-based alloys show distinctively more fragile behavior ($ D^{*}\sim 10-15 $). The most fragile is Pd$ _{43} $Ni$ _{10} $Cu$ _{27} $P$ _{20} $ that shows low temperature fragility ($ D^{*}=10.3 $) similar to that measured in molten state in Zr-based alloys [47, 56]. The result of the VFT-fitting for the glass formers Mg$_{65}$Cu$_{25}$Y$_{10}$ and the Zr$ _{46.75} $Ti$ _{8.25} $Cu$ _{7.5} $Ni$ _{10} $Be$ _{27.5} $ reproduced that of Refs. 36 and 1, respectively. For the Zr$ _{41.2} $Ti$ _{13.8} $Cu$ _{12.5} $Ni$ _{10.0} $Be$ _{22.5} $ only the 3-point beam bending data from Ref. 41 are considered for fitting, resulting in a $ D^{*} $ of 22, which was similarly found in Ref. 47. For the glass former Zr$ _{58.5} $Cu$ _{15.6} $Ni$ _{12.8} $Al$ _{10.3} $Nb$ _{2.8} $, the results are the same as that of Ref. 62. For the Cu$ _{47} $Ti$ _{34} $Zr$ _{11} $Ni$ _{8} $ only data for isothermal three-point beam bending in Ref. 78 were considered for fitting and yield $ D^{*} = 20.4 $. The data in Ref. 78 at high temperatures resulting from the non-contact oscillating drop technique are not shown in Fig. \ref{Fig16} because the temperatures lie above the strong into fragile transition temperature suggested by Evenson et al. [56] and show, indeed, when used in the VFT fitting, much smaller fragility parameter of about 12. The VFT-fitting on Pd$ _{40} $Ni$ _{40} $P$ _{20} $ yields a $ D^{*} $ parameter of 15.4 and $ T_{0} $ of 396 K. For the Pd$ _{40} $Cu$ _{30} $Ni$ _{10} $P$ _{20} $ glass former the fitting was performed using only five viscosity data points of Ref. 79 covering a temperature range from 635 K to 680 K. The fitting resulted in $ D^{*} = 14.5 $ and $ T_{0} = 418 $ K. For the Pd$ _{43} $Ni$ _{10} $Cu$ _{27} $P$ _{20} $ only the isothermal three-point beam bending values from Ref. 25 were taken into consideration for the VFT-fitting as well as the Adam-Gibbs fitting. Recent work of Gallino et al.[44] on the Pt$ _{57.3} $Cu$ _{14.6} $Ni$ _{5.3} $P$ _{22.8} $ showed a more fragile behavior for alloy than the one found in Ref. 18.  A VFT-fit of the relaxation times calculated from the DSC $ T_{g} $-shift due to different heating rates lead in Ref. 18 to an apparent stronger behavior for the Pt$ _{57.3} $Cu$ _{14.6} $Ni$ _{5.3} $P$ _{22.8} $, ($ D^{*}$ was about 16). This apparent stronger behavior was proved afterward Ref. 63 to reflect the cooling rate applied to the samples prior the DSC scans. Thus, the relaxation times taken from scans where the cooling rate is equal the heating rate would have been more appropriate in Ref. 18. In Fig. \ref{Fig16}, a fragility parameter of $ D^{*} = 12.5 $ and $T_{0}=357$ K is more representative for this alloy. These values where obtained in Ref. 44 by fitting the isothermal three-point beam bending viscosity data. 

Figure \ref{Fig18} shows the configurational entropy increase with undercooling resulting from the Adam-Gibbs fitting for each of the nine BMG formers selected for this study. There is a remarkable agreement between the Adam-Gibbs and VFT fits, reflected by the fact that the sequence of the curves in Fig. \ref{Fig18} follows the general trend observed in the VFT plot of Fig. \ref{Fig16}. This sequence goes from the stronger Zr-based formers to the more fragile Pt- and Pd-based formers, from left to right. For the Pt$ _{57.3} $Cu$ _{14.6} $Ni$ _{5.3} $P$ _{22.8} $, ($ D^{*}$, the recently obtained values [44] are $C=168.7$ kJ g-atom$ ^{-1} $,  $T_{m}^{*}=925$ K, $ T_{0}^{*}=375$ K and $S_{c}( T_{m}^{*})=18.92$ J g-atom$ ^{-1} $ K$ ^{-1} $.  The Pd$ _{43} $Ni$ _{10} $Cu$ _{27} $P$ _{20} $ is the most fragile liquid which is in agreement with the finding of Fig. \ref{Fig16}. For this alloy we have used the specific heat capacity data obtained in this study for fitting viscosity data to the Adam-Gibbs equation- due to the high consistency and the accuracy of the step method. In fact, there is excellent agreement between the experimentally determined enthalpies of crystallization for the Pd$ _{43} $Ni$ _{10} $Cu$ _{27} $P$ _{20} $ and the enthalpy function of the liquid that resulted from fitting the measured specific heat capacities (see Fig. \ref{Fig1}). This agreement verifies the accuracy of the thermodynamic experimental procedures of the 'step method' as well of $ T_{f} $ and $ \Delta H_{f} $ values used in Eq. (\ref{Eq12}), and substantiates the values of the thermodynamic constants $ a $, $ b $, $ c $ and $ d $ given above to describe the temperature dependence of $ c_{p} $. 

The Kauzmann (or isentropic) temperature of Pd$ _{43} $Ni$ _{10} $Cu$ _{27} $P$ _{20} $ is found in this study at 532 K. At this temperature the entropy of this multi-component glass former equals that of its crystalline counterpart. This temperature is about 100 K higher than the temperature where the configurational entropy of liquid is found to vanish, indicating that there is a considerable amount of entropy present in the four-component crystalline mixture. For each of the glass formers of this study, the configurational entropy is found to vanish at a temperature $ T_{0}^{*} $ that is below that of the corresponding Kauzmann temperature, and much closer to the VFT temperature $ T_{0} $. In all of the cases the isentropic temperature, does not seem to have a physical meaning. It has been already argued that in a multicomponent system the crystalline phases can have considerable configurational entropy resulting from the entropy of mixing [36,62]. Consequently, in a highly short range or medium range ordered liquid - like in the case of a deeply undercooled liquid - the entropy of the supercooled liquid could, in fact, become smaller than that of the crystalline mixture. Recently, Tanaka [80] reports $ T_{0} $ below $ T_{K} $ also for a number of non-metallic glass formers. There are also earlier examples where a crystalline metallic solid can have a higher entropy than the amorphous counterpart; for example for the case of inverse melting as reported by Vonallmen et al. [81] or Bormann et al.[82] In a one component system the Kauzmann temperature is indeed the lower bound for the glass transition since there is no contribution of the entropy of mixing with the exception of the small contribution due to vacancies in the crystal. However, multicomponent alloys have considerably increased configurational entropy due to presence of large amounts of solutes in the phases of the crystalline mixture. This has been experimentally confirmed for Zr-Cu-Ni-Al-Nb alloys where the crystalline mixtures consist of binary intermetallic compounds containing all the other elements as solutes [83]. In this regard, it is also worth noting that the calculated configurational entropy at $ T_{m}^{*} $ in this work is considerably larger than the ideal entropy of mixing for all of the alloys. This implies that these melts exhibit a considerable amount of excess entropy beyond the ideal mixture. In turn, the undercooled liquids of bulk metallic glass formers develop tremendous short and medium range order as recently been discussed by Miracle [84] and Ma et al.[85], which indicates that very low entropic states are possible for the deeply undercooled liquid. 

If we compare the obtained $ T_{m}^{*} $ value and the fit parameter $S_{c}( T_{m}^{*} )$, we find that they are somewhat larger that the melting temperature and the entropy of fusion, respectively. This is expected since $S_{c}( T_{m}^{*} )$ was chosen in the fitting procedure as the configurational entropy of a liquid above the melting point for a fixed liquid viscosity value of 1 Pa s, knowing that multicomponent BMG-forming liquids have high melt viscosities, usually larger than 1 Pa s. For example for the Pd$ _{43} $Ni$ _{10} $Cu$ _{27} $P$ _{20} $ BMG forming liquid the $S_{c}( T_{m}^{*} )$ is about 11 J g-atom$ ^{-1} $K$ ^{-1} $ at $ T_{m}^{*} $ of 900 K, whereas the entropy of fusion and the calorimetric melting peak temperature are about 6 J g-atom$ ^{-1} $K$ ^{-1} $ and 818 K, respectively. 

The free energy barrier for cooperative rearrangements $ C $ progressively increases with decreasing fragility. The stronger the glass, the larger is the $ C $. In Fig. \ref{Fig17}, $ C $ increases from about 150 kJ g-atom$ ^{-1} $K$ ^{-1} $ for the most fragile liquid to about 300 kJ g-atom$ ^{-1} $K$ ^{-1} $ for the strongest. This is certainly a reasonable range, since activation energies for diffusion around the glass transition temperature lie in that range; about 1 eV [circa 100 kJ g-atom$ ^{-1} $K$ ^{-1} $] for small atoms and up to 3 eV [circa 300 kJ g-atom$ ^{-1} $K$ ^{-1} $] for large atoms[75]. This result can be seen in the light of the discussion about cooperativity and correlation lengths. Intuitively in the strong liquid a flow or relaxation event is more localized (in a sense solid-like) and therefore an higher activation barrier needs to be overcome involving the large atoms. In Fig. \ref{Fig17}, a $D^{*}=26.5$ for Vitreloy 1 has been selected to represent the stronger liquid state for this alloy as described in the next paragraph. In Fig. \ref{Fig17},  only the Mg$ _{65} $Cu$ _{25} $Y$ _{10} $ BMG does not fit this general trend and shows a $ D^{*} $ value that is relatively high in comparison with its activation energy for cooperative rearrangements. The fact that the Mg-based alloy behaves differently than the Zr-, Ti- and noble metal-based alloys can also be seen when the fragility parameter, $ D^{*} $, is plotted against the number of components in the alloy, as it is recently been pointed out by Shadowspeaker and Busch [3]. They show that $ D^{*} $ increases monotonically as the complexity of the alloy increases and that only Mg$ _{65} $Cu$ _{25} $Y$ _{10} $ does not align well with the other alloy systems.

The results of Ref. 18 show also that $ T_{0} $, in the VFT fits, matches very well the temperature $ T_{0}^{*} $, where the configurational entropy vanishes in the Adam-Gibbs fits. When the configurational entropy of the liquid vanishes only one packing set is possible. This leads to the smallest possible potential energy of the system. The fact that the kinetic $ T_{0} $ and the thermodynamic $ T_{0}^{*} $ are similar, means that when the barrier with respect to viscous flow becomes infinitely large (at $ T_{0} $), the liquid will act like a solid that has assumed the ideal packing configuration. One might envision this configuration packing state similar to those proposed by Miracle [84] or by Ma et al.[85] derived from geometrical considerations. For a one component system this temperature $ T_{0} $ is also equal to the Kauzmann temperature, but as we see in the present study it is not true for a multi-component system.

\section{FRAGILE TO STRONG TRANSITIONS}
\label{sec:FS}

Recent studies into the melt viscosity of the Vitreloy 1 bulk metallic glass (BMG) have not only produced a picture of a highly viscous, dense metallic liquid, but have also revealed the presence of a distinct liquid-liquid transition within the equilibrium melt itself [2]. The characteristics of this fragile-to-strong transition can be understood within the context of the fragility scheme [5,6].

\begin{figure}
\begin{center}
\includegraphics[width=3in]{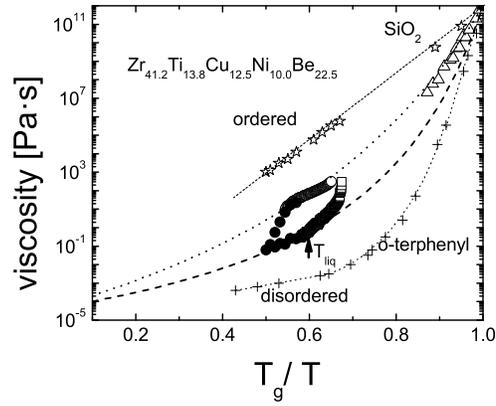}
\caption{Expanded Angell plot of combined isothermal ($ \bullet $) and viscosity measurements obtained at a constant clockwise shear rate while cooling at 2 K s$ ^{-1} $ from both 1125 K ($ \bigcirc $) and 1225 K ($ \Box $).  VFT fitting has been performed and shows the strong (...) and fragile (---) boundaries for Vitreloy 1.  The viscosity hysteresis shows the transition between these two boundaries with respect to temperature.  Also shown is TMA low viscosity measurements ($ \bigtriangleup $) obtained from three point beam bending experiments with an estimated shear rate of 10$ ^{-5} $ s$ ^{-1} $ [14].   This viscosity data matches the predicted strong VFT fits thus indicating a similar ordered state both below and above $ T_{liq} $ (taken from Ref. 47).}
\label{Fig19}
\vspace*{-12pt}\end{center}
\end{figure}

Experimental viscosity measurements on Vitreloy 1 in the vicinity of the glass transition temperature, $ T_{g} $, have determined a fragility parameter of $ D^{*} = 22 $, showing that this BMG is a moderately strong glass-former, similar to sodium silicate glasses [41]. However, it was shown that Vitreloy 1 retains its highly viscous, kinetically strong nature upon melting and, with increasing temperature, transforms to a more kinetically fragile system, characterized by a marked decrease in the viscosity of around three orders of magnitude [47]. The resulting viscosity as a function of  inverse temperature is shown in Fig. \ref{Fig19} as a fragility plot. The viscosity exhibits a pronounced hysteresis. This liquid remains in the fragile state until undercooled below the liquidus temperature, where the high viscosity liquid behavior is then re-established. According to Ref. [47], the fragility parameter of the strong Vitreloy 1 liquid, including the high-temperature data below the transition, is $ D^{*}= 26.5$, while the lower bound for the fragility parameter of the fragile liquid is $ D^{*} = 12$.  In addition to this fragile-to-strong transition, it was also shown in those same experiments that the Vitreloy 1 melt exhibits pronounced shear thinning behavior; both effects have been attributed to the destruction and re-establishment of short and medium-range order in the melt.

\begin{figure}
\begin{center}
\includegraphics[width=3in]{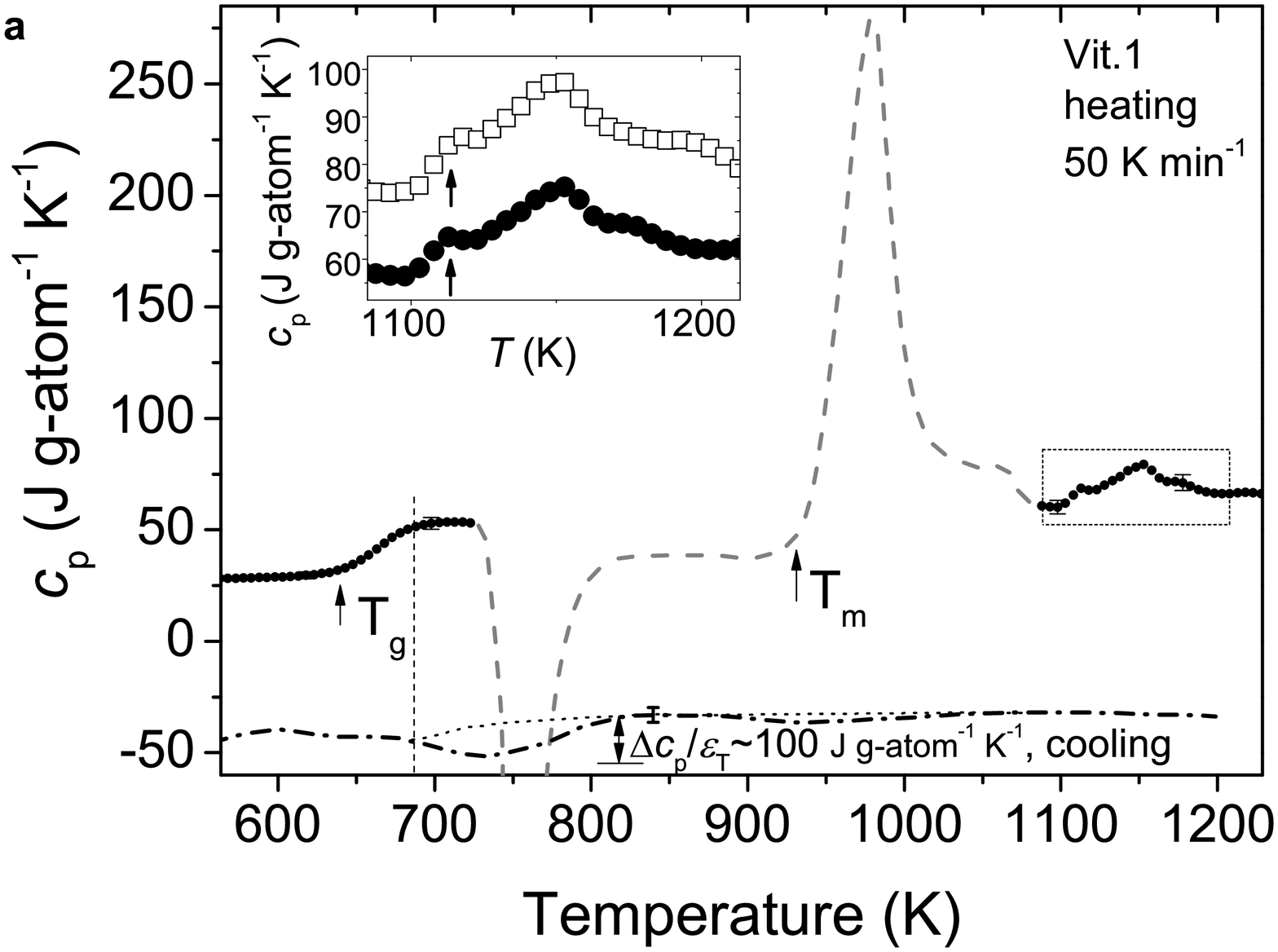}
\includegraphics[width=3in]{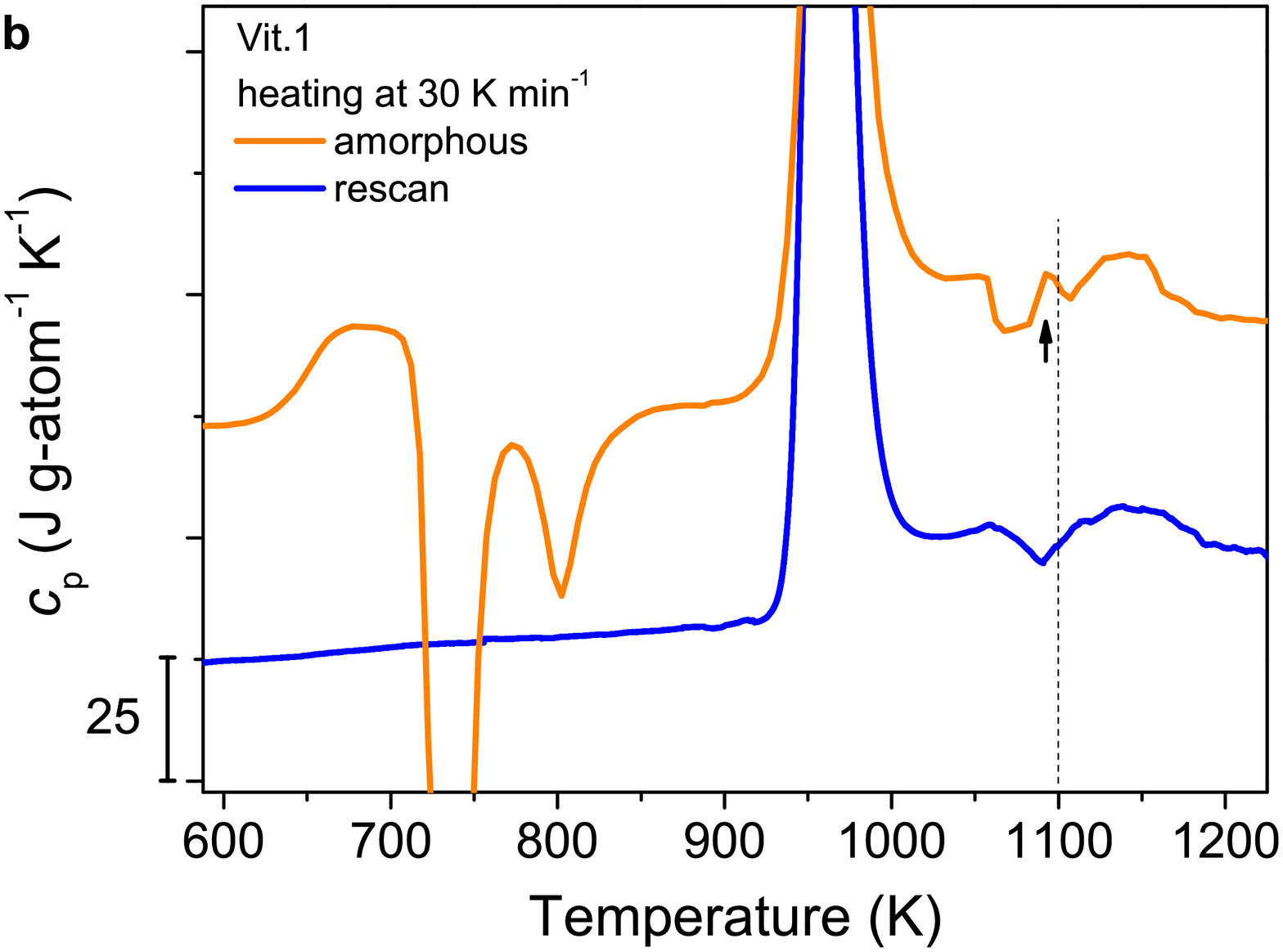}
\caption{Heat capacity $ c_{p} $ of Vitreloy 1. (\textbf{a}), $ c_{p} $ of an amorphous sample is measured upon heating at 50 K min$ ^{-1} $. Solid circles represent the glassy, supercooled liquid and stable liquid states; the dashed curve indicates the crystallization and melting processes. Note that there is a heat capacity peak at around 1100-1200 K, which occurs in the molten liquid according to the in-situ XRD taken at a 10 times higher heating rate ($ \sim $ 9 K s$ ^{-1} $) (see Fig. 2 of ref. [86]). Inset shows the magnification of the $ c_{p} $ peak (1100-1200 K) where the solid circles and open squares represent two separate measurements (vertically shifted for clarity). The arrows indicate that there is a small subpeak on the left shoulder of the main peak. In the lower part of the figure, the dash-dot curve shows the heat capacity during cooling of Vitreloy 1 taken from $ c_{p}/\epsilon T $ in ref. [75] (assuming the emissivity [75] $ \epsilon T=0.18 $), which is here plotted as negative values to indicate the exothermic event around 700-800 K in the supercooled liquid region in reference to the baseline (dotted curve) (see main text). (\textbf{b}), $ c_{p} $ measured upon heating at 30 K min$ ^{-1} $ for the amorphous sample (upper) and once-melted crystallized sample (lower) (vertically shifted for clarity). The arrow shows the small subpeak separated from the main peak to a lower temperature (reprinted from Ref. [86]). }
\label{Fig20}
\vspace*{-12pt}\end{center}
\end{figure}

Recently, Wei et al. [86] found the thermal signature of this fragile to strong transition. Figure \ref{Fig20} shows the heat capacity ($ c_{p} $) of Vitreloy 1 measured in reference to sapphire in graphite crucibles using calorimetry. A heat capacity peak is observed on heating between around 1100 K and 1200 K, above the reported liquidus temperature 1026 K. The area of the $ c_{p} $ peak is proportional to the heat gain, which is determined to be $ \Delta H_{LL}\approx 1.0\pm 0.1 $ kJ g-atom$ ^{-1} $, about 10\% of the enthalpy of fusion ($ \Delta H_{f}\approx 9.7\pm 0.7 $ kJ g-atom$ ^{-1} $). The inset shows the zoom-in of the peak (solid circles) and a separate scan (open squares) in which the main peak is reproduced (1100-1200 K) [86].  We notice that a small subpeak on the left shoulder of the broad peak is also reproducible. By lowering the heating rate down to 30 K min$ ^{-1} $, this small subpeak can be separated to a lower temperature $ <1100 $ K from the main broad peak (see the upper curve in Fig. \ref{Fig20}b). And during a rescan of the once-melted crystallized sample, the subpeak disappears while the main peak remains (lower curve in Fig. \ref{Fig20}b). This observation suggests that this small subpeak probably comes from a small portion of remaining crystalline phases. The first scan apparently reduces the inhomogeneity and thus diminishes the small subpeak and ruggedness of the measured heat capacity curve. However, the broad main peak ($ \sim $1100-1200 K) cannot be explained by melting of crystals according to the results of in-situ synchrotron X-ray diffraction experiments and volume measurements [86]. This $ c_{p} $ peak should be considered as a consequence of an intrinsic change in the liquid. The dash-dot line in the lower part of Fig. \ref{Fig20} is the deep supercooled $ c_{p} $ data from $ c_{p} /emissivity$ extracted from a temperature-time profile measured in an ESL by Ohsaka et al. [75]. An exothermic peak is reported around 700-800 K corresponding to an enthalpy release $ \sim $900 J g-atom$ ^{-1} $ estimated by the authors, which, by the following analysis alternative to the authors', should be associated with the $ c_{p} $ peak on heating (1100-1200 K) when taking the viscosity hysteresis and structural measurements into account [86]. 

Both $ c_{p} $ peaks on heating and cooling correspond to a similar enthalpy change ($ \sim $1 kJ g-atom$ ^{-1} $) and form a hysteresis with respect to temperature, which is comparable to the viscosity hysteresis that characterizes the strong-fragile crossover (see Fig. \ref{Fig19}) [86]. The correlation between thermodynamics and kinetics is suggested by Adam-Gibbs theory [17], at least, qualitatively. These hysteresis phenomena are consistent with the hysteresis-like behaviour in liquid structural changes observed in the first peak position and full width at half maximum (FWHM) of the structure factor $ S(Q) $ using synchrotron X-ray scattering (Fig. 3 b, c of Ref. 86) where two different local structures were found corresponding to two liquid states with distinct properties.

These hysteresis phenomena suggest that there exists a reversible weak first-order liquid-liquid transition between two liquid phases with different entropy, fragility and local structures in Vitreloy 1 system, in which the high temperature liquid needs to be supercooled and the low temperature liquid needs to be overheated to nucleate the respective other liquid. The authors proposed a homogeneous nucleation scenario for the mechanism of the suggested liquid-liquid transition [86]. On heating, the fragile droplet nucleates homogeneously in the strong liquid matrix at $ \sim $1100 K. On cooling the reversible transition occurs at $ \sim $830 K through the homogeneous nucleation of strong liquid droplets in the fragile liquid matrix. Apparently, there is considerable undercooling and overheating involved, where a faster cooling rate may cause a lower transition temperature. Thus, the structural transition during cooling at a rate of $ \sim $ 10 K s$ ^{-1} $ in electrostatic levitator are detected at a somewhat lower temperature ($ \sim $800 K) than the kinetic fragile-to-strong transition ($ \sim $900 K) observed at a cooling rate of $ \sim $2 K s$ ^{-1} $ in the viscosity measurements. According to the classic nucleation theory, the homogeneous nucleation rate depends on both the diffusion (viscosity) and the energy barrier for the critical nucleus, $ \Delta G^{*}\propto \gamma_{LL}^{3}/\Delta T_{c}^{2} $, where $ \gamma_{LL}\propto \Delta S_{LL} $ is the fragile/strong liquid interfacial energy, $ \Delta S_{LL} $ the entropy difference across the interfaces and $ \Delta T_{c} $ is either the critical undercooling $ \Delta T_{c}^{u} $ or overheating $ \Delta T_{c}^{o} $. In this scenario, if the critical temperature $ T_{c} $ for the first-order liquid-liquid transition ($ \Delta G_{LL}=0 $) is assumed to be located approximately in the middle between 830 K and 1100 K at 965 K (with $\Delta T_{c}^{u}=\Delta T_{c}^{o}=135 $ K), the entropy difference between the strong and fragile liquid can be estimated as $ \Delta S_{LL}\approx 1 $ J g-atom$ ^{-1} $ K$ ^{-1} $. This makes it at first surprising that large undercooling (overheating) is necessary to overcome this barrier. However, it needs to be emphasized here that Vitreloy 1 exhibits a very sluggish liquid kinetics, which results in both slow nucleation and growth kinetics of the respective other liquid phase even with a rather small barrier for homogeneous nucleation. One fundamental difference compared to a first-order liquid-crystalline transition has to be pointed out. During melting of a crystalline solid the liquid forms spontaneously at internal interfaces, such as grain boundaries and at the surface by heterogeneous nucleation of the melt and virtually no overheating is observed. In contrast, in the strong liquid that transforms into the fragile liquid, few internal interfaces exist and heterogeneous nucleation is rare, leading to overheating in this case [86].

A liquid-liquid transition and a phase separation can occur simultaneously when the composition of the system is not right on the critical composition for a pure polyamorphic transition. In such a case, the high-temperature fragile phase separates into two phases: one is strong and the other remains fragile. A slight compositional change does not affect the intrinsic liquid structural change which is the real origin of the drastic viscosity and fragility change of the liquid. This is analogous to the face-centered-cubic Cu-Au system where the first-order order-disorder transition and the phase separation occur simultaneously with decreasing temperature when the composition of Cu-Au is slightly off the critical ratio, 3:1.

\begin{figure}[h]
\begin{center}
\includegraphics[width=3in]{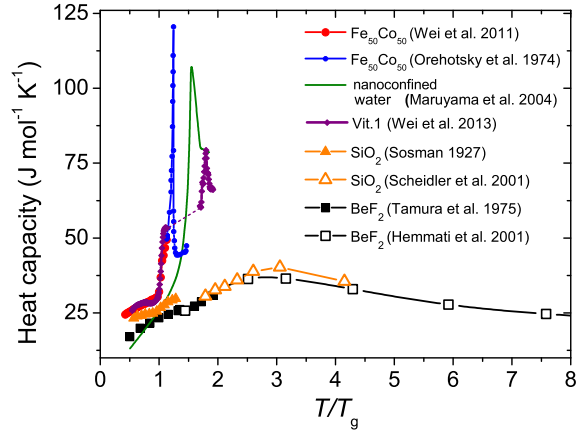}
\caption{Comparison of heat capacity maxima between the liquid Vitreloy 1 and other glass formers. Solid and open symbols represent experimental and simulation data, respectively (for detailed data sources, see Ref. [86] and [87]). The heat capacity values are plotted against the $ T_{g} $-scaled temperature. For both SiO$ _{2} $ and BeF$ _{2} $, the heat capacity maxima with the dynamic crossover are located beyond the normal measurement range, far above $ T_{m} $, suggesting that the liquid-liquid transitions are in the stable liquid state at the high temperature. In the case
of water, the suggested liquid-liquid transition is in the supercooled liquid regime where the $ c_{p} $ peak is observed in the water confined by nanopores to avoid crystallization. The liquid-liquid transition of Vitreloy 1 upon heating is above $ T_{m} $ and has a sharper $ c_{p} $ peak than that of SiO$ _{2} $ and BeF$ _{2} $. These liquid-liquid transitions are considered as off-critical phenomena, comparing with the critical phenomenon of the lambda (order-disorder) transition in the non-liquid superlattice Fe$ _{50} $Co$ _{50} $ with a very sharp lambda $ c_{p} $ peak (reproduced from Ref. 86 and 87). }
\label{Fig21}
\vspace*{-12pt}\end{center}
\end{figure}

\begin{figure}[h]
\begin{center}
\includegraphics[width=3in]{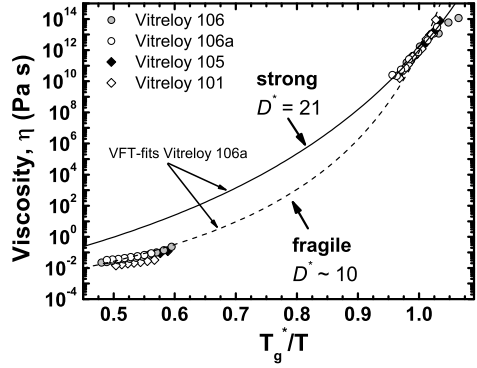}
\caption{Angell plot over the entire temperature range of measured viscosities for the alloys  Vitreloy 106, Vitreloy 106a, Vitreloy 101 and Vitreloy 105. At low temperatures, near $ T_{g} $, the isothermal equilibrium viscosities were measured using three point beam bending. Separate VFT fits to the low and high temperature viscosity data of, for example, Vitreloy 106a are included here as solid and dashed curves, respectively. A VFT fit of the low temperature three point beam bending viscosity data yields a fragility parameter $ D^{*} of 21 $ [40]. From fitting the Vitreloy 106a viscosity data taken at temperatures above $ T_{liq}$, $D^{*} $ was determined to be $ \sim 10 $ (taken from Ref. 56).}
\label{Fig22}
\vspace*{-12pt}\end{center}
\end{figure}

Liquid Vitreloy 1 apparently fits into the strong class of Angell's fragility pattern [6] and is comparable to the archetypical strong liquids, SiO$ _{2} $ and BeF$ _{2} $ as well as water that are involved in a liquid-liquid transition. Saika-Voivod et al. [88] revealed a fragile-to-strong transition of liquid SiO$ _{2} $ associated with a heat capacity anomaly above the melting temperature $ T_{m} $ by studying static and dynamic properties of liquid silica using numerical simulations. Molten BeF$ _{2} $ studied by Hemmati et al. [89] using the ion dynamics simulations exhibits, also, a fragile-to-strong crossover correlated with a heat capacity maximum above $ T_{m} $. Oguni et al. [90] and Chen et al. [91] confined supercooled water within silica gel nanopores to avoid crystallization. A pronounced heat capacity peak is observed at about 225 K above $ T_{g} $ and below $ T_{m} $ and this peak is accompanied with a fragile-to-strong transition evidenced by a number of studies [91,92].

For comparison, $ c_{p} $ vs. $ T_{g} $-scaled temperature for Vitreloy 1, SiO$ _{2} $, BeF$ _{2} $ and nanoconfined water are plotted in Fig. \ref{Fig21}, also with the $ c_{p} $ of a non-liquid superlattice system Fe$ _{50} $Co$ _{50} $ which has a glass transition (kinetic freezing-in) during the lambda (order-disorder) transition [87]. These substances with the anomalous $ c_{p} $ peaks resemble a system with a lambda (order-disorder) transition that is driven into off-critical behaviour, for example, by increasing the pressure [87]. A liquid-liquid transition can be understood as an underlying lambda (order-disorder) transition that separate a strong liquid below the transition temperature from a fragile liquid above it. As a consequence, strong liquids are expected to experience such a transition above $ T_{g} $, which can be observed when the observation window is appropriate and crystallization is avoided. Recent viscosity measurements (Fig. \ref{Fig22}) suggest that a liquid-liquid transition may also exist in the Zr-based bulk metallic glass-formers, Vitreloy 106, Vitreloy 106a, Vitreloy 101 and Vitreloy 105 [56].

\section{CONCLUSIONS}

Since bulk metallic glass formation became a wide spread phenomenon in the 1990$ ^{th} $, supercooled metallic liquids became accessible in a much wider temperature and time window than before. They are relative easy and straightforward to investigate. Glass transition, crystallization and melting are experimentally clearly defined. The crystalline ground state is accessible. The time-temperature-transformation diagrams are found to be on the laboratory temperature and time scale. Therefore, thermodynamics, viscosity and relaxation phenomena can be determined quantitatively and analyzed using the common models on thermodynamics and kinetics.

Bulk metallic glass forming liquids are densely packed and as a consequence they show high viscosity and sluggish crystallization kinetics when compared with other metallic liquids. If compared with other substances they behave intermediate between the strongest and the most fragile liquids. BMG can be described in the frameworks of the free volume model as well as the Adam-Gibbs theory with volume, enthalpy and viscosity relaxing on the same time scale. Especially the analysis with the Adam Gibbs equation shows that the magnitude of the  excess specific heat capacity in the supercooled liquid and thus the change in entropy is quantitatively correlated to the kinetic fragility expressed by the empirical VFT equation. In BMG the entropy of fusion is not a measure for the difference in configurational entropy between liquid and crystal. The crystalline equilibrium phases have high entropies due to entropy of mixing. In turn the supercooled liquids can exhibit pronounced short and medium range order. Therefore the temperature where the configurational entropy vanishes in the Adam-Gibbs model and the temperature where the barrier to flow diverge in the VFT description are found considerably lower than the isentropic (Kauzmann) temperature for all bulk metallic glasses. Those temperatures, obtained independently with both approaches match closely to within 5-10\%.

There appear to exist at least two families of BMG forming liquids. Nobel metal based metallic glasses with relatively low affinity to oxygen are relatively fragile. Their crystallization is nucleation controlled. Fluxing with B$ _{2} $O$ _{3} $ and thus cleaning them, makes them robust with respect to crystallization. Alloys that combine predominantly early with late transition metals are more affine to oxygen and can not be fluxed. However they tend to be stronger and therefore have a more sluggish kinetics. Both nucleation and especially growth is retarded in these alloy. It appears that there might be a third family of alloys that are based on elements like Mg or Ca that are extremely sensitive to oxygen. Apparently, they are even stronger and crystallization is only growth controlled since nucleation can not be avoided altogether. Here more data have to be collected.

A polyamorphic liquid-liquid is observed in Vitreloy 1, which appears to be a transition from a fragile liquid at high temperatures to a strong liquid at low temperatures. It must be a weak first order phase transformation analogous to disorder-order transition in simple fcc crystal structures. If the composition of the melt is not right on the critical point of the polyamorphic transformation, cooling will involve phase separation into at least one fragile and one strong liquid.



%
%

%

\section*{ACKNOWLEDGMENTS}
The authors thank the agencies that have funded this work over the past two decades. These include the Alexander von Humboldt Foundation, the US Department of Energy, the National Aeronautics and Space Administration, the Defense Advanced Research Projects Agency, the National Science Foundation, the Deutsche Forschunggemeinschaft (German Research Foundation) and the German Federal Ministery for Economics and Technology. We also thank Liquidmetal Technology for providing Zr-based alloys and C. Hafner Precious Metal Technology for providing the noble metals. Furthermore we would like to thank W.L. Johnson, J. Schroers, A. Masuhr, T.A. Waniuk, C. Way, P. Wadhwa, M. Stolpe as well as C.A. Angell for their contributions to this work. This paper was presented (published in a special volume) at the Symposium on Fragility, Bangalore, dedicated to Austen Angell for his 80th birthday, 2014. 


\end{document}